\documentclass[aps,prd,twocolumn,superscriptaddress]{revtex4-2}

\usepackage{graphicx}
\usepackage{amsmath}
\usepackage{amssymb}
\usepackage{siunitx}
\usepackage{hyperref}
\hypersetup{
	colorlinks = true,
	linkcolor = blue,
	citecolor = blue,
	urlcolor  = blue,
}
\usepackage{color}
\usepackage[normalem]{ulem}

\begin{document}

\title{Density of states and differential entropy  in Dirac materials in  crossed magnetic and in-plane electric fields}

\author{Andrii~A.~\surname{Chaika}}
\affiliation{Department of Physics, Taras Shevchenko National University of Kyiv,
64/13, Volodymyrska Street, Kyiv 01601, Ukraine}

\author{Yelizaveta \surname{Kulynych}}
\affiliation{Department of Physics, The \'{E}cole polytechnique f\'{e}d\'{e}rale de Lausanne, Lausanne 1015, Switzerland}

\author{D.~O.~\surname{Oriekhov}}
\affiliation{Kavli Institute of Nanoscience, Delft University of Technology, 2628 CJ Delft, the Netherlands}

\author{Sergei~G.~\surname{Sharapov}}
\affiliation{Bogolyubov Institute for Theoretical Physics, National Academy of Sciences of Ukraine,
Kyiv 03143, Ukraine}
\affiliation{
Kyiv Academic University, Kyiv 03142, Ukraine}

\date{\today}

\begin{abstract}
The density of states and differential entropy per particle are analyzed for Dirac-like electrons in
graphene subjected to a perpendicular magnetic field and an in-plane electric field. For comparison,
the derived density of states is contrasted with the well-known case of nonrelativistic electrons in
crossed magnetic and electric fields.
The study considers ballistic electrons and also includes the effect of small impurity scattering.
In the latter case, the limit of zero magnetic field and the so-called collapse of Landau levels in graphene
are examined analytically. By comparing the results with numerical calculations on graphene ribbons, we demonstrate that the Landau state
counting procedure must be modified for Dirac-like electrons, leading to a fields-dependent 
Landau level degeneracy factor.
Additionally, it is shown that peaks in the differential entropy arise from the dispersionless surface mode
localized at the zigzag edges of the ribbon.
\end{abstract}


\maketitle

\section{Introduction}

The entropy $S$ of many-body systems plays a fundamental role in characterizing their thermodynamic behavior,
heat transfer, and thermoelectric properties.
On the other hand, directly measuring entropy experimentally has always been challenging
(see, however, Ref.~\cite{Adam2024}).
Recently, however, it has been found that the differential entropy, $s \equiv \partial S/\partial n$, where 
$S$ is the entropy per unit volume and 
$n$ is the electron density, can be investigated experimentally.
Nevertheless, the experiment described in \cite{Kuntsevich2015NatComm}
is not an exception, as the quantity measured directly is actually the temperature derivative of the chemical potential, $\partial \mu/\partial T$.
These derivatives are then equated using
the Maxwell relation:
\begin{equation}
\label{entropy-part}
s = \left( \frac{\partial S}{\partial  n} \right)_T = -\left( \frac{\partial \mu }{\partial T} \right)_n.
\end{equation}
The differential entropy $s$ serves as an excellent thermodynamic tool, demonstrating high sensitivity in low charge density regimes,
which allows for the study of two-dimensional electron gas (2DEG) in gated structures.
In particular, in Ref.~\cite{Kuntsevich2015NatComm} the measurements of $s$ were performed in a
quantizing magnetic field $H$ perpendicular to 2DEG. The entropy
showed a nonmonotonic dip-peak behavior as a function of $H$.  As discussed in
\cite{Blanter1994PR} (see also a recent review \cite{Varlamov2021LTP}) an
intersection of the sequent Landau level and the chemical potential level
can be viewed as an example of the Lifshitz electronic topological transition (ETT).
Furthermore, measurements of differential entropy and integrated entropy change have revealed the isospin
Pomeranchuk effect in magic-angle twisted bilayer graphene, as reported in
Refs.~\cite{Saito2021Nature,Rozen2021Nature}.
An important property of entropy per particle is its close relation to more complex transport properties, such as the Seebeck coefficient, making it a useful indicator for predicting the thermoelectric behavior of materials \cite{Cortes2023PRB} (see also Refs.~\cite{Varlamov2016PRB,Tsaran2017SciRep}).

Theoretical results for $s(\mu)$ indicate that it displays peak-dip structures associated
with ETT   in various 2D materials, such as gapped graphene monolayers \cite{Tsaran2017SciRep},
germanene \cite{Grassano2018PRB}, and
semiconducting dichalcogenides \cite{Shubnyi2018FNT}.
It is shown in \cite{Kulynych2022}
that in the specific case of ETT occurring when the chemical potential crosses the saddle point in the dispersion, the
differential entropy can be used to identify the type of associated van Hove singularity.

For an infinite system, the energies of the relativistic Landau levels in graphene subjected to a perpendicular magnetic field $H$ and an in-plane electric field $E$ are given by \cite{Lukose2007PRL,Peres2007JPCM}
\begin{equation}
\label{LL-collapse}
\mathcal{E}_{n,k}  = \mathcal{E}_{n}  - \hbar k \frac{c E}{H}, \quad
\mathcal{E}_{n}= \pm  (1 - \beta^2 )^{3/4} \sqrt{2 n} E_M ,
\end{equation}
where $n=0,1, \ldots$,
$k$ is the in-plane wave vector along the direction perpendicular to the electric field,
$E_M = \hbar v_F/l$ is the magnetic energy scale with
$l = \sqrt{\hbar c /(e H)}$ being the magnetic length,
$\beta  =  v_0/v_F = c E/(v_F H)$, $v_F$ is the Fermi velocity and $v_0$ is the
drift velocity. Here and in what follows we assume
that $H > 0$ and use CGS units.

As the dimensionless parameter $\beta$, which characterizes the strength of the electric field for a given magnetic field, approaches its critical value $|\beta_c| = 1$, the Landau level staircase collapses into a single level \cite{Lukose2007PRL,Peres2007JPCM}. In an infinite system, this collapse can be interpreted as a transition from closed elliptic quasiparticle orbits when $|\beta| < 1$ ($|v_0| < v_F$) to open hyperbolic orbits when
$|\beta| > 1$ ($|v_0| > v_F$) \cite{Shytov2009SSC}.

It is worth noting an interesting connection between the problem of collapsing Landau levels and studies involving tilted Dirac cones in
a strong magnetic field \cite{Goerbig2009EPL}. However, the key distinction lies in the term $\hbar k c E /H$, which lifts the Landau level degeneracy
but does not induce the tilt.

Note also that there is no Landau level collapse in ribbons because the orbit center cannot extend to infinity. Instead, the electron- and hole-like
levels on opposite edges of the ribbon become denser, with the distance between them scaling as $O(l^3/W^2)$, where $W$ is the ribbon
width \cite{Herasymchuk2024PRB}. The presence of disorder inevitably causes the broadening of Landau levels, and as a result, these levels will merge due to their finite width.

It was shown in \cite{Mahan1996PNAS} that a delta-shaped transport distribution function maximizes
the thermoelectric properties of a material.
This result indicates a narrow distribution of the
energy of the electrons participating in the transport.
Consequently, it is reasonable to expect that the thermoelectric properties of Dirac materials
will be enhanced as they approach the regime of Landau level collapse.

The aim of this study is to examine the behavior of entropy per particle $s$ in graphene subjected to crossed magnetic and in-plane electric fields,
with a focus on how level convergence influences it.
Since the calculation of $s$ relies on the density of states (DOS), the majority of this work is dedicated to the calculation and analysis of the DOS.
The paper is organized as follows.  In Sec.~\ref{sec:DOS-2DEG},
we provide an overview of the results for the DOS in crossed fields for ballistic electrons
in a nonrelativistic 2DEG \cite{Kubisa1997PRB, Zawadzki1999APP}. The notions of magnetic and electric
regimes are introduced.
An analytical expression for the DOS that is convenient for calculations and  takes into account quasiparticle
scattering is suggested. In Sec.~\ref{sec:DOS-Dirac}, the DOS for Dirac-like fermions is derived
and analyzed using both analytical and numerical methods, with a discussion on the Landau level degeneracy factor.
In Sec.~\ref{sec:entropy}, we demonstrate how the entropy per particle varies with the chemical potential as the ratio
of electric to magnetic fields changes. This analysis is carried out both for an analytical model, which considers only
the bulk Landau levels, and for numerical simulations on a ribbon, which include the dispersionless surface mode localized
at the zigzag edges of the ribbon.
Finally, the conclusions are given in Sec.~\ref{sec:concl}.

\section{DOS in the crossed fields: nonrelativistic case}
\label{sec:DOS-2DEG}

\subsection{Nonrelativisitic spectrum and general definition of the DOS}
First of all, we recapitulate the DOS behavior 
in crossed fields for ballistic electrons
in a nonrelativistic 2DEG \cite{Kubisa1997PRB, Zawadzki1999APP}. The spectrum of nonrelativistic 2D electrons
in a magnetic field $H$, perpendicular to the 2DEG plane, and an in-plane electric field $E$, directed along
the $x$-axis, is given by \cite{Galitski2013book}:
\begin{equation}
\label{spectrum-crossed-Schrodinger}
\begin{split}
\mathcal{E}_{n, k_y}^{\mathrm{NR}} & = \mathcal{E}_n^{\mathrm{L}} - \frac{m c^2 E^2}{2 H^2} - \hbar k_y \frac{c E}{H}, \\
\mathcal{E}_n^{\mathrm{L}} & = E_M^{\mathrm{NR}} \left(n+ \frac{1}{2} \right),
\end{split}
\end{equation}
where as in Eq.~(\ref{LL-collapse})  $n=0,1, \ldots$,
$k_y$ is the in-plane wave vector along $y$ direction, and
$E_M^{\mathrm{NR}} = \hbar \omega_c  $ is
the nonrelativistic magnetic energy scale with $\omega_c = e H/(m c)$
being the cyclotron frequency and $m$ the effective carrier
mass, respectively. The spin splitting is omitted both in
Eq.~(\ref{spectrum-crossed-Schrodinger}) and above in Eq.~(\ref{LL-collapse}).

The full or integrated DOS per spin and unit area is defined as the sum over the complete set of quantum numbers
$\alpha = (n, k_y)$, which  reads
\begin{equation}
\label{DOS-def}
D(\mathcal{E}) = \frac{1}{\mathcal{A}} \sum_{\alpha} \delta(\mathcal{E} - \mathcal{E}_\alpha) =
\frac{L_y}{\mathcal{A}} \int \frac{d k_y}{2 \pi} \sum_n  \delta(\mathcal{E} - \mathcal{E}_\alpha).
\end{equation}
We consider a system with dimensions $L_x$ and $L_y$, giving it an area of $\mathcal{A} = L_x L_y$.

In the absence of an electric field, $E=0$, the spectrum $\mathcal{E}_{n, k_y}^{\mathrm{NR}} $
simplifies to the standard Landau's spectrum, $\mathcal{E}_n^{\mathrm{L}}$ given by
the second line in  Eq.~(\ref{spectrum-crossed-Schrodinger}).
Exploiting the position-wave vector duality of Landau states, where the wave vector $k_y$ determines
the center of the electron orbital along the $x$-axis, given by $x_0 = -k_y l^2$, the DOS can be rewritten as follows:
\begin{equation}
\label{DOS-E=0-x0}
D_0^{\mathrm{NR}}(\mathcal{E}) = \frac{L_y}{\mathcal{A}} \int \frac{d x_0}{2 \pi l^2}  \sum_n  \delta(\mathcal{E} - \mathcal{E}_n^{\mathrm{L}}).
\end{equation}
Being independent of the quantum number $k_y$ the Landau levels are infinitely degenerate. As a consequence
of this, the DOS   (\ref{DOS-E=0-x0}) is ill-defined.
The procedure for regularizing the number of states involves counting only the states within $-L_x/2 \leq x_0 \leq L_x/2$
and then taking the limit $L_x \to \infty$:
\begin{equation} \label{limit}
\lim_{L_x \to \infty} \frac{1}{L_x}\int dx_0 \to \lim_{L_x \to \infty} \frac{1}{L_x}\int_{-L_x/2}^{L_x/2} dx_0 = 1 .
\end{equation}
This procedure is known as Landau state counting \cite{Landau1930ZP} (see also the textbook \cite{Hajdu.book}),
and it yields the conventional expression for the DOS:
\begin{equation}
\label{DOS-E=0}
D_0^{\mathrm{NR}}(\mathcal{E}) =  \frac{1}{2 \pi l^2} \sum_n  \delta(\mathcal{E} - \mathcal{E}_n^{\mathrm{L}}).
\end{equation}
Here
\begin{equation}
\label{degeneracy-LL}
\mathfrak{g}_L = \frac{1}{2 \pi l^2} = \frac{eH}{2 \pi \hbar c}
\end{equation}
is the usual degeneracy factor that corresponds to the number of states in the momentum space per one Landau level.
Quasiclassically, it is defined as:
\begin{equation} \label{degeneracy-def}
\mathfrak{g} = \frac{1}{(2 \pi \hbar)^2} \int \int d p_x d p_y= \frac{S_{n+1} - S_n}{(2 \pi \hbar)^2},
\end{equation}
where the integration is done over neighboring classical trajectories and
$S_n$ represents the orbit area in momentum space corresponding to the $n$-th Landau level.
Evidently, the degeneracy (\ref{degeneracy-LL}) is recovered by taking
$S_n = 2 \pi m \mathcal{E}_n^{\mathrm{L}}$.

Although DOS (\ref{DOS-E=0}) is derived by placing the electrons in a finite-size system, the Landau spectrum
$\mathcal{E}_n^{\mathrm{L}} $ in Eq.~(\ref{spectrum-crossed-Schrodinger}) corresponds to an infinite system
and does not take into account the surface electrons.
Nevertheless, there is a consistency
between the results obtained from Landau's theory \cite{Landau1930ZP}, essentially based on the DOS (\ref{DOS-E=0}),
and Teller's approach \cite{Teller1931ZP}, which accounts for the finite size of the system and the presence of the surface magnetization currents,
as discussed in Ref.~\cite{Hajdu1974ZP}.

\subsection{DOS in the crossed fields in the absence of scattering}

In the presence of an electric field, the center of the electron orbit shifts as \cite{Galitski2013book}
$x_0 = - k_y l^2 - m c^2 E/(e H^2)$. Accordingly, the spectrum (\ref{spectrum-crossed-Schrodinger})
in crossed fields can be expressed as follows:
\begin{equation}
\label{spectrum-crossed-Schrodinger-x0}
\mathcal{E}_{n, x_0}^{\mathrm{NR}} = \mathcal{E}_{n}^{\mathrm{NR}} + e E x_0, \quad
\mathcal{E}_{n}^{\mathrm{NR}} = \mathcal{E}_{n}^{\mathrm{L}} + \Delta, \quad \Delta = \frac{m c^2 E^2}{2 H^2} \,.
\end{equation}

Assuming that the orbit center falls within the range $-L_x/2 \leq x_0 \leq L_x/2$, we
obtain the following DOS in crossed fields \cite{Kubisa1997PRB, Zawadzki1999APP}:
\begin{equation}
\label{DOS-2DEG-Gamma=0}
\begin{split}
&D^{\mathrm{NR}}(\mathcal{E})  = \frac{L_y}{\mathcal{A}} \int_{-L_x/2}^{L_x/2} d x_0 D_0^{\mathrm{NR}} (\mathcal{E} - \Delta - e E x_0)= \\
&  \frac{\mathfrak{g}_L}{ U}
\sum_n \left[\theta(\mathcal{E} - \mathcal{E}_{n}^{\mathrm{NR}}+U/2) - \theta (\mathcal{E} - \mathcal{E}_{n}^{\mathrm{NR}} - U/2) \right].
\end{split}
\end{equation}
Here, the DOS defined by Eq.~(\ref{DOS-E=0}) is now written
for the spectrum $\mathcal{E}_{n}^{\mathrm{NR}}$ given by Eq.~(\ref{spectrum-crossed-Schrodinger-x0})
as taken into by the argument shift, $D_0^{\mathrm{NR}}(\mathcal{E} - \Delta)$,
and $U = e E L_x$ is the potential difference between $x=-L_x/2$ and $x = L_x/2$.
Here and in what follows, we assume for definiteness that $U >0$.
Clearly, in the limit of a vanishing electric field, $U \to 0$, Eq.~(\ref{DOS-2DEG-Gamma=0})
reduces to its derivative with respect to $U$, returning to Eq.~(\ref{DOS-E=0}).

Note that in the presence of an electric field, the final result depends on the choice of integration limits for $x_0$,
namely $-L_x/2 \leq x_0 \leq L_x/2$ or $0 \leq x_0 \leq L_x$, which leads to a shift in the levels' energies due to
the electric field \cite{Kubisa1997PRB, Zawadzki1999APP, Alisultanov2014JETPL}.
For clarity in the presentation, we leave out this shift, by using symmetric integration limits for $x_0$.

\subsection{Effect of elastic scattering}

Elastic scattering of electrons by defects and impurities is inevitably present in real systems and results in level
broadening. A simple way to account for this smearing is by introducing a finite electron lifetime $\tau$.
This broadens the Dirac delta function peaks associated with the Landau levels into
Lorentzians with a constant, energy-independent width $\Gamma = \hbar/\tau$, as follows (see Ref.~\cite{Sharapov2004PRB} for a detailed discussion):
\begin{equation}
\label{Lorentzian}
\delta(\mathcal{E} - \mathcal{E}_n) \to \frac{1}{\pi} \frac{\Gamma}{(\mathcal{E} - \mathcal{E}_n)^2 + \Gamma^2}.
\end{equation}
Accordingly, for numerical computations based on Eq.~(\ref{DOS-2DEG-Gamma=0}), one can replace the Heaviside theta function $\theta(\mathcal{E}_{n})$
in the DOS (\ref{DOS-2DEG-Gamma=0}) with
\begin{equation}
\label{theta-Gamma}
\tilde \theta (\mathcal{E}) = \frac{1}{2} + \frac{1}{\pi}\arctan \frac{\mathcal{E}}{\Gamma}.
\end{equation}

The Lorentzian approximation (\ref{Lorentzian}) allows us to derive, from Eq.~(\ref{DOS-E=0}), a simple
analytical expression for the DOS per spin and unit area in the absence of an electric field \cite{Slobodeniuk2011PRB}:
\begin{equation}
\label{DOS-E=0-Gamma}
\begin{split}
D_0^{\mathrm{NR}}(\mathcal{E}) = -\frac{m}{2 \pi^2 \hbar} \mbox{Im} \psi
\left( \frac{1}{2} - \frac{\mathcal{E} + i \Gamma}{E_M^{\mathrm{NR}}} \right) \\
= \frac{\mathfrak{g}_L}{\pi} \frac{d}{d \mathcal{E}}
\mbox{Im} \ln \Gamma \left(\frac{1}{2} - \frac{\mathcal{E} + i \Gamma}{E_M^{\mathrm{NR}}} \right),
\end{split}
\end{equation}
where $\psi(z)$ and $\Gamma(z)$ are the digamma and gamma functions,
respectively.
It is clear that the peaks (oscillations) in the DOS are embedded
in these functions when the real part of the argument becomes negative.

For a finite $\Gamma$  substituting Eq.~(\ref{DOS-E=0-Gamma}) into the first line of
Eq.~(\ref{DOS-2DEG-Gamma=0}), we obtain the DOS per unit area
\begin{equation}
\label{DOS-2DEG-Gamma}
\begin{split}
D^{\mathrm{NR}} (\mathcal{E})  =
\frac{\mathfrak{g}_L}{\pi  U}
&\mbox{Im} \left[ \ln \Gamma \left(\frac{1}{2} - \frac{\mathcal{E} -\Delta  + U/2 +i \Gamma}{E_M^{\mathrm{NR}}} \right) \right. \\
-& \left.
\ln \Gamma \left(\frac{1}{2} - \frac{\mathcal{E} -\Delta- U/2 + i \Gamma}{E_M^{\mathrm{NR}}} \right)
\right].
\end{split}
\end{equation}

\subsection{Illustrations of magnetic and electrical regimes}
\label{sec:illustrations-nonrel}

\begin{figure}[h]
   \centering
\includegraphics[width=\columnwidth]{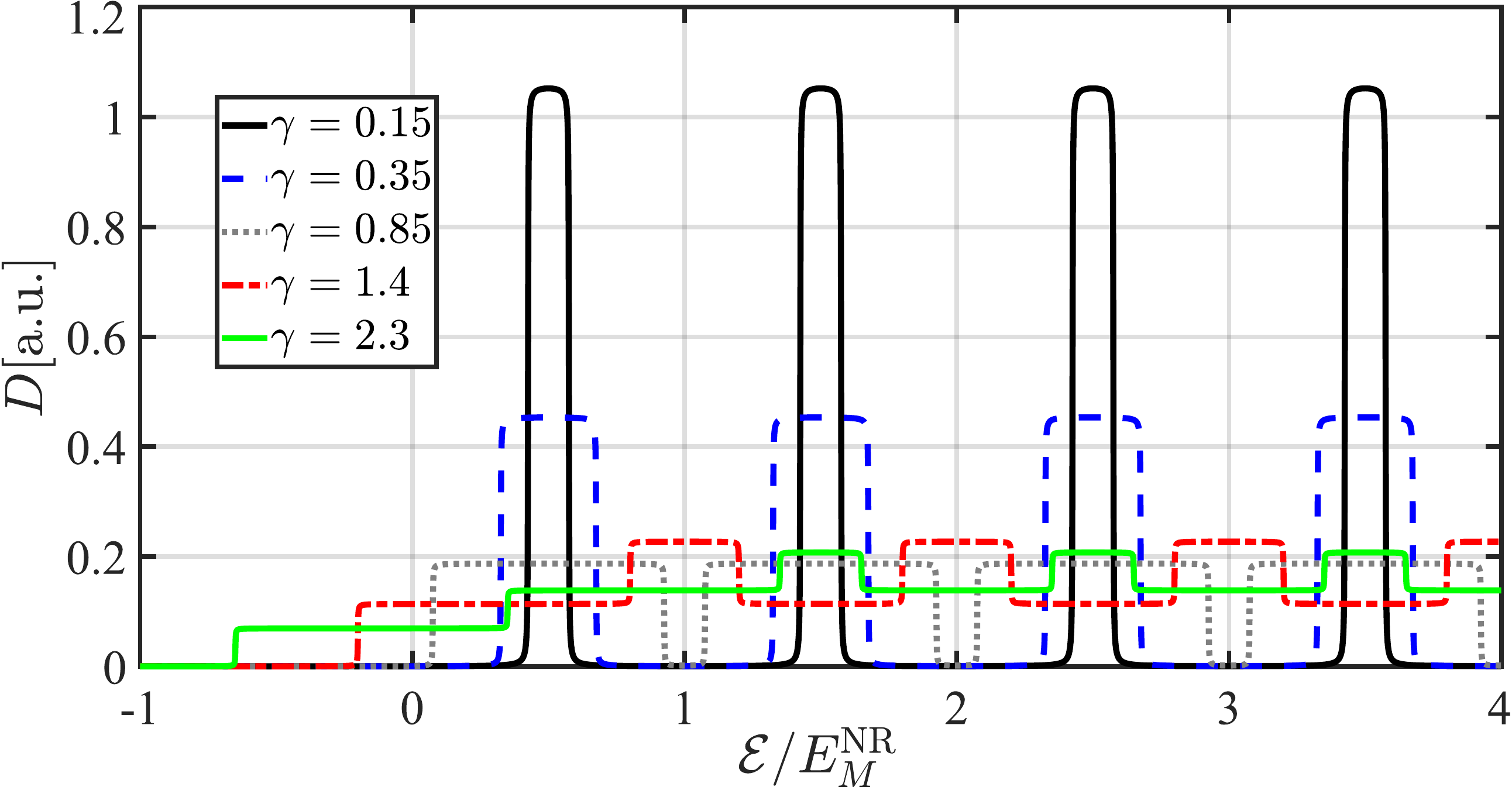}%
\caption{The DOS, $D^{\mathrm{NR}}(\mathcal{E})$, in crossed magnetic and electric fields
versus energy $\mathcal{E}$ in units of $E_M^{\mathrm{NR}} = \hbar \omega_c$ for
five values $\gamma = U /E_M^{\mathrm{NR}}$. The scattering rate $\Gamma=0.001 E_M^{\mathrm{NR}}$. }  \label{fig:1}
\end{figure}

\begin{figure}[h]
   \centering
\includegraphics[width=\columnwidth]{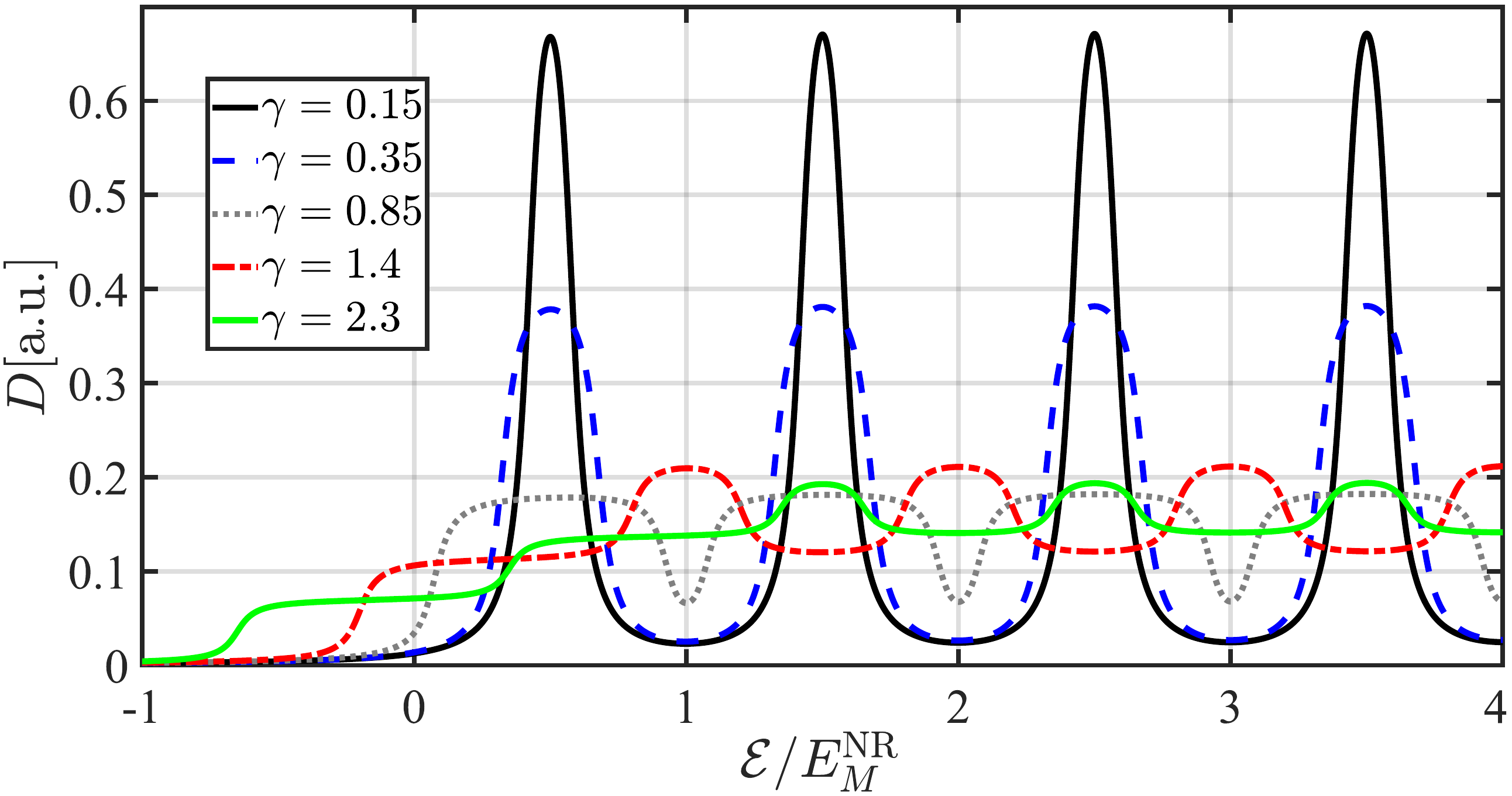}%
\caption{The DOS, $D^{\mathrm{NR}}(\mathcal{E})$, in crossed magnetic and electric fields
versus energy $\mathcal{E}$ in units of $E_M^{\mathrm{NR}}$ for the same
five values $\gamma = U /E_M^{\mathrm{NR}}$ as in Fig.~\ref{fig:1}.
The scattering rate  $\Gamma=0.05 E_M^{\mathrm{NR}}$. }
\label{fig:2}
\end{figure}

The characteristic cases of DOS (\ref{DOS-2DEG-Gamma=0}) in crossed fields are presented in
Figs.~\ref{fig:1} and \ref{fig:2} for the scattering rate $\Gamma =0.001 E_M^{\mathrm{NR}}$ and
$\Gamma=0.05 E_M^{\mathrm{NR}}$, respectively.
The linear shift of the levels due to the electric field is absent
because symmetric integration limits were chosen in Eq.~(\ref{DOS-2DEG-Gamma=0}).
The quadratic shift in the electric field by $\Delta$ present in $\epsilon_n$
[see Eq.~(\ref{spectrum-crossed-Schrodinger-x0})] is also
omitted as in \cite{Zawadzki1999APP}.
It can be seen that the presence of an electric field destroys the
sharp ($\delta$-like in the absence of scattering) peaks in the DOS.
This occurs due to the $k_y$ dependence of the energies in Eq.~(\ref{spectrum-crossed-Schrodinger}).
The parameter that controls the behavior of the DOS is the ratio of electric to magnetic energy,
$\gamma = U/E_M^{\mathrm{NR}} = e E L_x/(\hbar \omega_c)$.
Accordingly, the regimes $\gamma < 1$ and $\gamma > 1$ are referred to as the magnetic and electric regimes,
respectively \cite{Kubisa1997PRB, Zawadzki1999APP}.
For $\gamma \ll 1$, the peaks in the DOS are well-defined, although their width is not related
to the scattering rate $\Gamma$. For $\gamma \sim 1$, the peaks become lower and close to overlapping,
and for $\gamma > 1$, contributions from different subbands associated with the different
Landau levels overlap significantly.

As mentioned above, plotting Figs.~\ref{fig:1} and \ref{fig:2}
we omitted the shift of $\sim E^2/H^2$ of the energies.  It follows from the energies $\mathcal{E}_{n}^{\mathrm{NR}}$
in Eq.~(\ref{spectrum-crossed-Schrodinger-x0}) that the conduction subbands shift
{\it upwards \/} as $E/H$  increases.
On the other hand, using the equivalent form of the spectrum (\ref{spectrum-crossed-Schrodinger})
which contains $- m c^2 E^2/(2 H^2) - \hbar k_y c E /H$ and integrating
over $k_{min} \leq k_y  \leq k_{max}$ with $k_{min} = - k_{max}$, we find that
the remaining conduction subband shifts {\it downwards \/} when $E/H$
increases.
Thus, depending on whether the $k_y$ dependence of the subbands is included
in $x_0$ as  in Eq.~(\ref{spectrum-crossed-Schrodinger-x0}) or not, they
shift upwards or downwards, at the rate $m c^2 E^2/(2 H^2)$, respectively \cite{Kubisa1997PRB, Zawadzki1999APP}.
As discussed in Refs.~\cite{Kubisa1997PRB, Zawadzki1999APP},
in the case of direct optical transitions, the value $k_y$ is conserved.
This results in the downward subband shift, which was observed experimentally.
However, it is important to remember that in real samples, the broadening of DOS peaks due
to the $k_y$ dependence of the energies is several orders of magnitude larger than the shift,
allowing the latter to be neglected.
As we will see later, the entire Landau state-counting procedure requires revision in the case of the
collapsing Dirac spectrum (\ref{LL-collapse}).
This issue will be revisited in the next section.

\subsection{DOS in the limit of vanishing magnetic field}

An advantage of the analytic representations (\ref{DOS-E=0-Gamma}) and
(\ref{DOS-2DEG-Gamma}) is that it allows access to the limit
of the vanishing magnetic field, $H \to 0$. Using
the asymptotic expansions
\begin{equation}
\label{digamma-expand}
\psi(z) = \ln z - \frac{1}{2z} - \frac{1}{12 z^2} + O \left(\frac{1}{z} \right)^4, \qquad  z \to \infty,
\end{equation}
and
\begin{equation}
\label{ln-Gamma-expand}
\begin{split}
\ln \Gamma(z)  = & \left(z - \frac{1}{2} \right) \ln z - z + \frac{1}{2} \ln (2 \pi) \\
& + \frac{1}{12z} - \frac{1}{360z^3} + O \left(\frac{1}{z} \right)^4, \qquad  z \to \infty,
\end{split}
\end{equation}
and also taking the $\Gamma \to 0$ limit,
one obtains, respectively, the DOS
\begin{equation}
\label{DOS-2DEG-free}
D_0^{\mathrm{NR}} (\mathcal{E})  =
\begin{cases}
0, & \mathcal{E} < 0,\\
\frac{m}{2 \pi \hbar^2} , & \mathcal{E} >0,
\end{cases}
\end{equation}
and
\begin{equation}
\label{DOS-2DEG-electric}
D^{\mathrm{NR}} (\mathcal{E}) =
\begin{cases}
0, & \mathcal{E} \leq -U/2,\\
\frac{m}{2 \pi \hbar^2} \frac{U + 2 \mathcal{E}}{2U}, & -U/2 \leq \epsilon \leq U/2,\\
\frac{m}{2 \pi \hbar^2} , & \mathcal{E} \leq U/2.
\end{cases}
\end{equation}
Eq.~(\ref{DOS-2DEG-free}) is nothing but the free electron DOS per spin in 2D that confirms
the consistency of Landau state counting.
The result  (\ref{DOS-2DEG-electric})
is also consistent with the corresponding expressions from Refs.~\cite{Kubisa1997PRB, Zawadzki1999APP},
where it was derived for the case of a zero magnetic field and a finite electric field.

\section{DOS in the crossed fields: case of graphene}
\label{sec:DOS-Dirac}

\subsection{DOS in the crossed fields in the absence of scattering}

The key difference between the spectrum (\ref{LL-collapse}) and the 2DEG spectrum (\ref{spectrum-crossed-Schrodinger})
is that the Dirac Landau levels are not equidistant.
Consequently, as we will see below, the parameter $\gamma = e E L_x / E_M$, with
\begin{equation}
\label{magnetic-energy}
E_M =  \hbar v_F/l \equiv \sqrt{ \hbar v_F^2 e H / c},
\end{equation}
which characterizes the ratio of electric to magnetic energies, cannot globally distinguish between
the magnetic and electric regimes across all energy levels.

Using Eq.~(\ref{degeneracy-def}) with $S_n = \pi \mathcal{E}_{n}^2/v_F^2$, we find that the factor $\mathfrak{g}$
becomes dependent on $\beta$ and is expressed as:
\begin{equation}
\label{degeneracy-Dirac}
\mathfrak{g} (\beta) = \mathfrak{g}_L (1 - \beta^2)^{3/2}.
\end{equation}
This dependence was not accounted for in previous work on graphene in the crossed fields \cite{Alisultanov2014JETPL, Alisultanov2014PB}.
As we will demonstrate, using the Landau degeneracy factor $\mathfrak{g}_L$ instead of $\mathfrak{g}(\beta)$ leads
to unphysical behavior in the DOS. In what follows we will show that our choice of $\mathfrak{g}(\beta)$ is consistent both
with numerical calculations performed on ribbons and with the zero field DOS.
We also note that a related study \cite{Goerbig2008PRB} (see also Ref.~\cite{Goerbig2009EPL}) on tilted Dirac cones in a magnetic field
demonstrated that calculating the DOS requires accounting for the renormalization of the effective Fermi velocity, averaged along a
semiclassical elliptical trajectory.

Thus by substituting the Dirac spectrum (\ref{LL-collapse}) into the definition (\ref{DOS-def}),
integrating over the wave number $-L_x/(2 l^2) \leq k_y \leq L_x/(2 l^2) $,
and taking into account the proposed degeneracy factor (\ref{degeneracy-Dirac}),
we obtain the DOS for both valleys per spin:
\begin{equation}
\label{DOS-Dirac}
\begin{split}
 D (\mathcal{E}) =
\frac{2\mathfrak{g} (\beta)}{U}
\sum_n & \left[\theta(\mathcal{E} -  \mathcal{E}_{n}+U/2) \right. \\
& - \left. \theta (\mathcal{E} -  \mathcal{E}_{n} - U/2) \right],
\end{split}
\end{equation}
where as above $U = e E L_x$.
Note that the method of integrating over wave numbers $k_y$ used here corresponds to the procedure described
at the end of Sec.~\ref{sec:illustrations-nonrel}. An alternative approach could involve integrating over
the orbit center position $x_0$, which, in the case of Landau level collapse in an infinite system,
is given by \cite{Lukose2007PRL, Sari2015PRB, Herasymchuk2024PRB}:
\begin{equation} \label{orbital-infinite}
x_0 = - k_y l^2 - \frac{\beta l \, \text{sgn} (\mathcal{E}_n) \sqrt{2n}}{(1 - \beta^2)^{1/4}}.
\end{equation}
The different choices of integration limits
reflect the difficulties with
the Landau state counting procedure, especially in the presence of electric field.
In such cases, the geometric parameter $L_x$ enters the final result
through the electric potential difference $U$.
As mentioned above, the procedure adopted here ensures agreement with numerical results obtained
for sufficiently wide ribbons, where the portion of the spectrum within the ribbon aligns with the spectrum
(\ref{LL-collapse}) for an infinite system.

It is helpful to rewrite the DOS (\ref{DOS-Dirac}) in terms of the dimensionless parameter $\gamma$ as follows:
\begin{equation}
\label{DOS-Dirac-dimensionless}
\begin{split}
D (\epsilon) =
\frac{2\mathfrak{g} (\beta)}{ E_M \gamma}
\sum_n & \left[\theta(\epsilon -  \epsilon_{n}(\beta) +\gamma/2) \right. \\
& - \left. \theta (\epsilon -  \epsilon_{n}(\beta) - \gamma/2) \right],
\end{split}
\end{equation}
where the energy is measured in the units of $E_M$
\begin{equation}
\label{LL-collapse-dimensionless}
\epsilon_n (\beta) \equiv \mathcal{E}_{n}/E_M = \pm  (1 - \beta^2 )^{3/4} \sqrt{2 n}, \quad n =0,1, \ldots
\end{equation}

By expressing the electric energy as $U = \beta \hbar v_F L_x/ l^2$, we see that the parameters $\gamma$ and $\beta$
are related by
\begin{equation}
\label{gamma-via-beta}
\gamma = \beta L_x / l.
\end{equation}
Thus, varying $\beta$ leads to changes in $\gamma$. It is more convenient to consider both parameters as independent variables.
This can be achieved by assuming that, for fixed values of $\beta$, variations in $\gamma$ are obtained by adjusting the width $L_x$.
Similarly, one can assume that $L_x$ is adjusted to keep $\gamma$ constant as $\beta$ changes.

\subsection{Peculiarities of magnetic and electric regimes in the Dirac Landau level case}
\label{sec:Dirac-illustr}

To illustrate the role of $\gamma$ in the Dirac Landau level case, it is convenient to start by setting $\beta = 0$
in Eq.~(\ref{LL-collapse-dimensionless}) and examining how the DOS behaves as the value of $\gamma$ varies in Eq.~(\ref{DOS-Dirac-dimensionless}).
Recall that level broadening can be incorporated by replacing $\theta(\epsilon)$ with $\tilde{\theta}(\epsilon)$,
as defined by Eq.~(\ref{theta-Gamma}).

\begin{figure}[h]
   \centering
\includegraphics[width=\columnwidth]{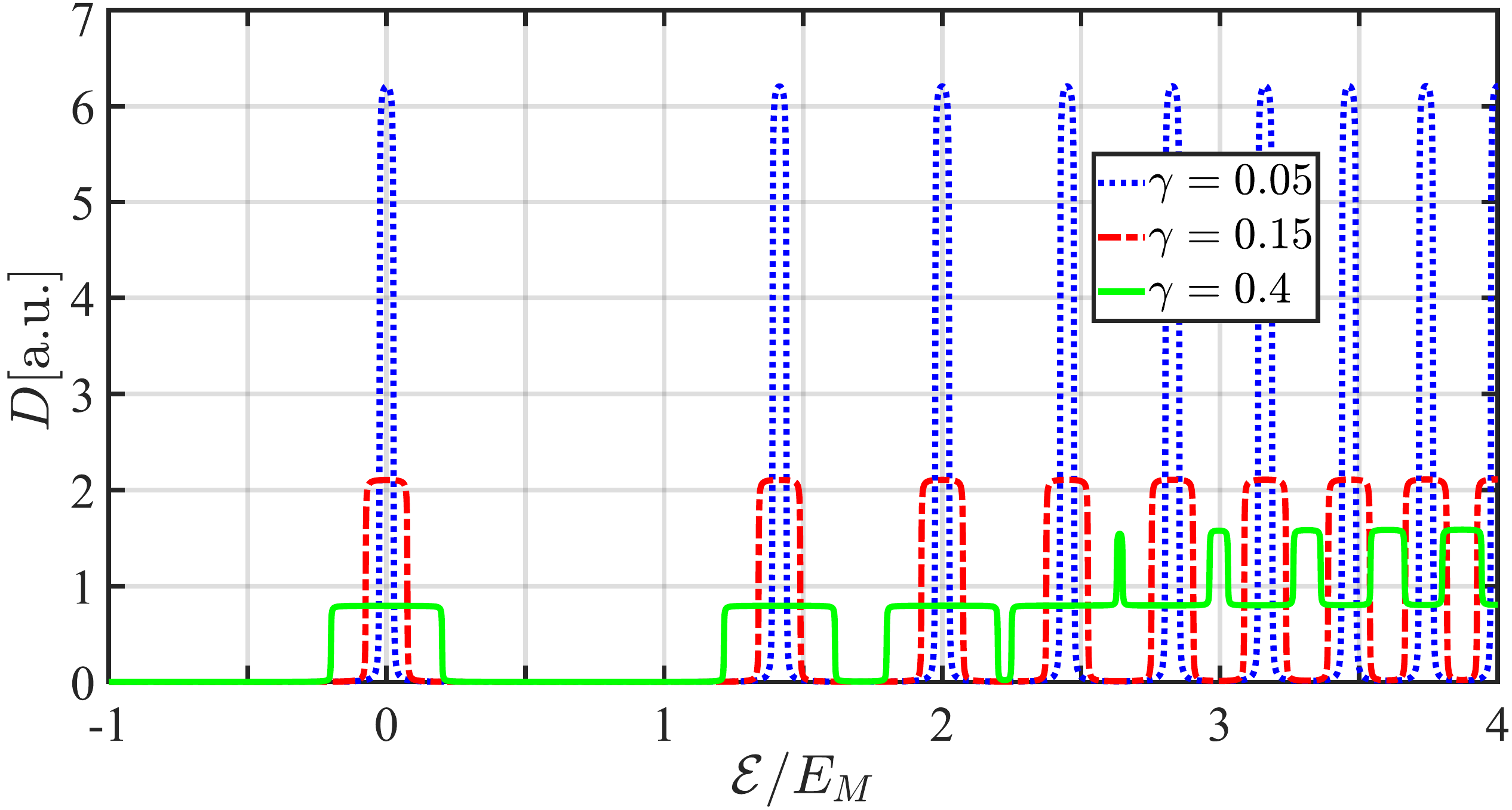}%
\caption{The DOS, $D(\mathcal{E})$, in crossed magnetic and electric fields
versus energy $\mathcal{E}$ in units of $E_M$ for three values $\gamma = U /E_M$.
The scattering rate $\Gamma=0.001 E_M$.}
\label{fig:3}
\end{figure}

In Fig.~\ref{fig:3}, we present the DOS (\ref{DOS-Dirac-dimensionless}) for $-1 \leq \epsilon \leq 4$.
Note that with symmetric integration limits,  the DOS remains an even function of $\epsilon$,
as in the zero electric field case. Thus, here and in what follows, we plot the DOS over a representative energy range.
The curves for smaller values, $\gamma = 0.05$ and $\gamma = 0.15$, within the shown energy range, correspond
to the magnetic regime. The peaks' centers align with the positions of the Landau levels, given by $\epsilon(\beta = 0) = \pm \sqrt{2 n}$.
The widening of the levels is due to the $k_y$ dependence of the level energies, rather than the Landau level width $\Gamma = 0.001 E_M$, which is
chosen to be rather small.
For the larger value of $\gamma = 0.4$, levels with $n > 2$ begin to overlap, indicating that the system has entered the electric regime.

In general, the condition for the overlap of levels $\mathcal{E}_{n+1}$ and $\mathcal{E}_{n}$ is given by:
\begin{equation}
\label{overlap}
\mathcal{E}_{n+1} - \mathcal{E}_{n} = U .
\end{equation}
This leads to the condition
\begin{equation}
\sqrt{n+1} + \sqrt{n} = \frac{\sqrt{2} (1-\beta^2)^{3/4}}{\gamma}
\end{equation}
which, for $n \gg 1$, simplifies to
\begin{equation}
\label{overlap-approx}
n \simeq \frac{(1-\beta^2)^{3/2}}{2 \gamma^2} .
\end{equation}
For $\gamma = 0.4$, the last equation yields $n \simeq 3.1$,
which we observe despite this being a relatively low value for the level index $n$.

\begin{figure}[h]
   \centering
\includegraphics[width=\columnwidth]{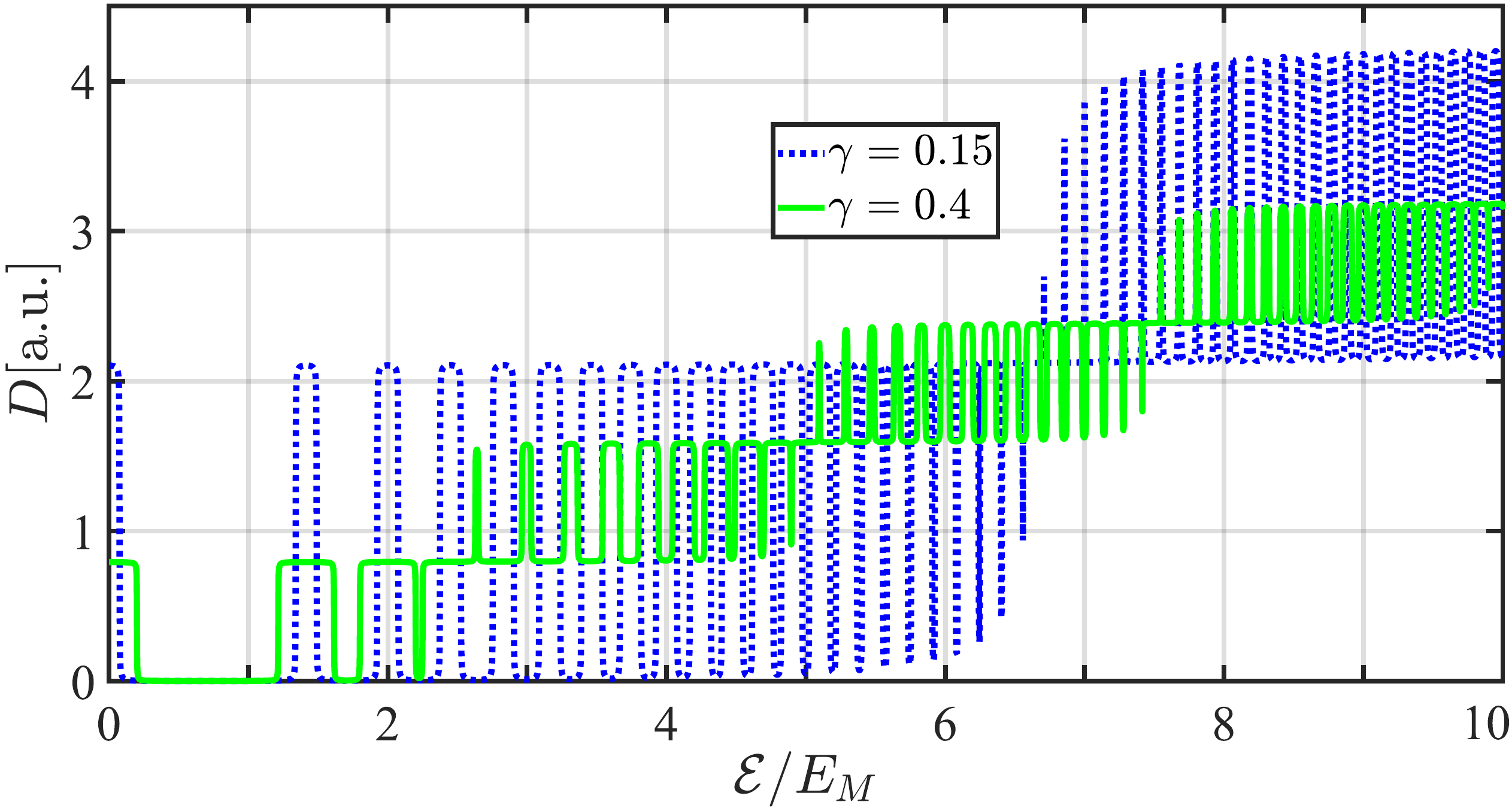}%
\caption{The DOS, $D(\mathcal{E})$, in crossed magnetic and electric fields
versus energy $\mathcal{E}$ in units of $E_M$ for two values $\gamma = U /E_M$.
The scattering rate $\Gamma=0.001 E_M$.}  \label{fig:4}
\end{figure}

\begin{figure}[h]
   \centering
\includegraphics[width=\columnwidth]{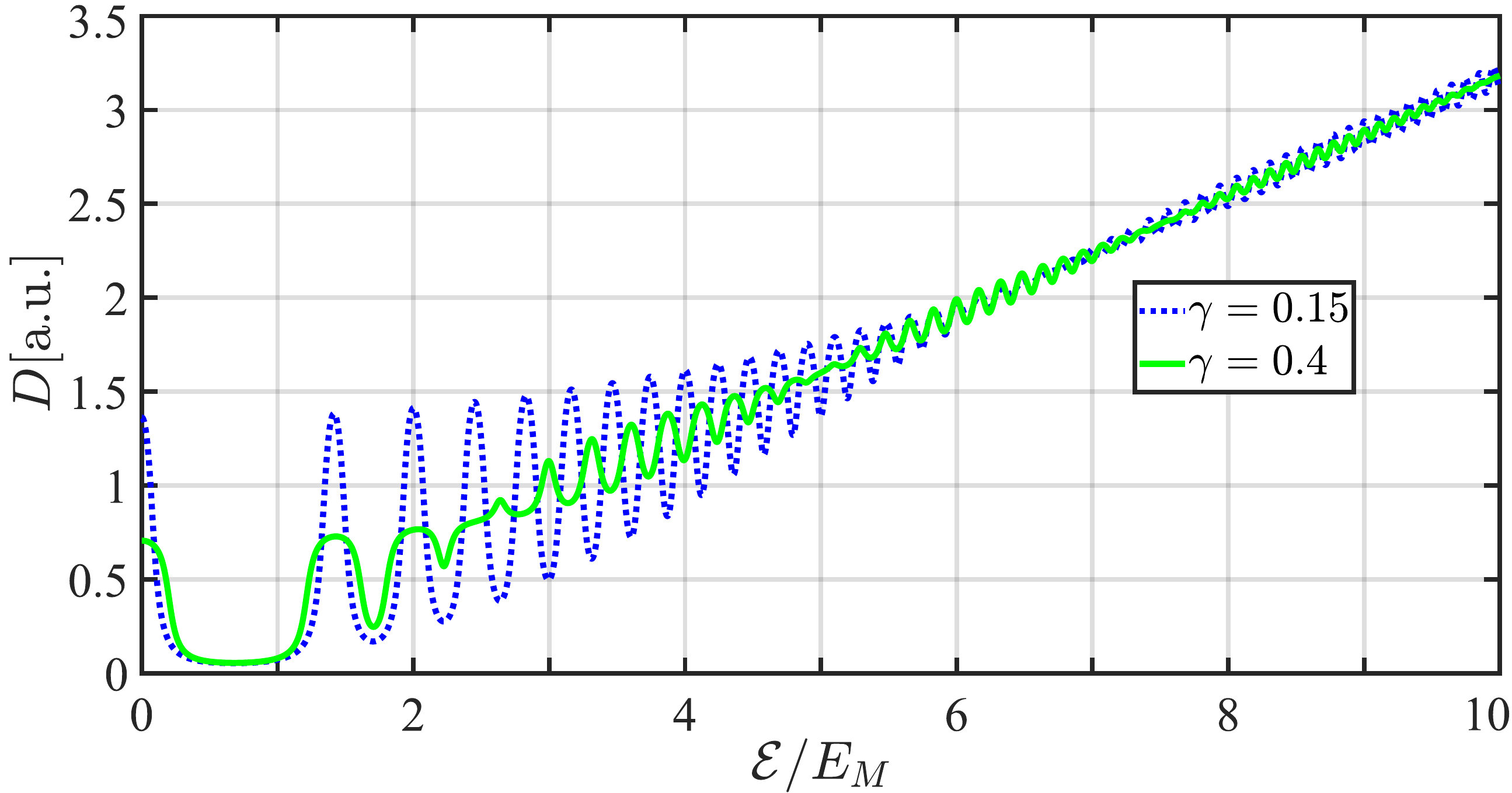}%
\caption{The DOS, $D(\mathcal{E})$, in crossed magnetic and electric fields versus energy $\mathcal{E}$, in units of $E_M$,  for
the same two values of $\gamma = U /E_M$ as in Fig.~\ref{fig:4}. The scattering rate
$\Gamma=0.05 E_{M}$.}  \label{fig:5}
\end{figure}

\begin{figure}[h]
   \centering
\includegraphics[width=1\columnwidth]{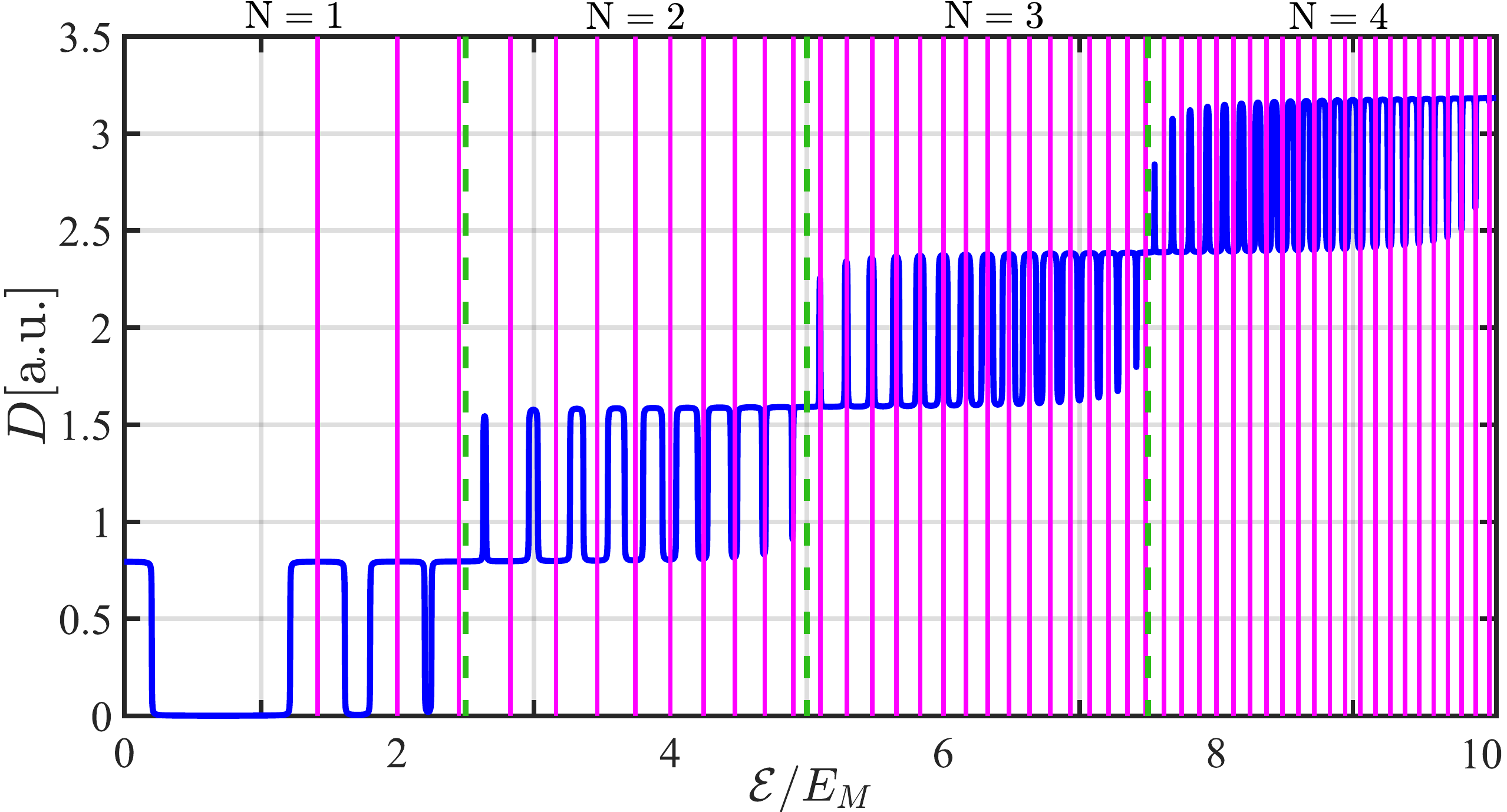}%
\caption{The DOS, $D(\mathcal{E})$, in crossed magnetic and electric fields versus energy $\mathcal{E}$ in units of $E_M$  for
$\gamma = U /E_M=0.4$. The scattering rate $\Gamma=0.001 E_{M}$. The solid
(magenta) vertical lines show the positions of the Landau levels $\epsilon_n (\beta=0)$ given by Eq.~(\ref{LL-collapse-dimensionless}).
The dashed (green) vertical lines demarcate the boundaries of the regions with the different number
of overlapping Landau levels.}  \label{fig:6}
\end{figure}

In Fig.~\ref{fig:4} we show the DOS (\ref{DOS-Dirac-dimensionless}) for the same two values of $\gamma=0.15$ and  $\gamma = 0.4$  as
above in Fig.~\ref{fig:3}, but for a wider range of energies $0 \leq \epsilon \leq 10$. One can see that while in Fig.~\ref{fig:3}
the red curve
corresponding to $\gamma = 0.15$ remained in the magnetic regime, in Fig.~\ref{fig:4} the same curve (blue)
now enters electric regime
for $n \gtrsim 22$ in agreement with Eq.~(\ref{overlap-approx}).
Furthermore, the curve for $\gamma = 0.4$, which enters the electric regime even at low energies $\epsilon \gtrsim 2.5$,
exhibits an interesting pattern as the energy increases. We observe a step-like increase in the DOS, with each step containing
oscillations of the same amplitude. As we will discuss in relation to Fig.~\ref{fig:6},
this behavior is associated with the overlap of three or more Landau levels.
Fig.~\ref{fig:5} is calculated for the same values of $\gamma$ and the same range of energies as  Fig.~\ref{fig:4},
but with a Landau level width $\Gamma$ that is 50 times larger. This results in the smearing of the DOS
for larger values of $\epsilon$, and as we will see below, its behavior resembles the DOS in the absence of the
magnetic field.

We now return to the DOS envelope steps observed in Fig.~\ref{fig:4}. In Fig.~\ref{fig:6}
we again plot the DOS for $\gamma = 0.4$, but with additional vertical lines to clarify the observed behavior.
The solid (magenta) lines correspond to the positions of the Landau levels unperturbed by electric
field  given by Eq.~(\ref{LL-collapse-dimensionless}) for $\beta =0$.
We recall that Figs.~\ref{fig:3} -- \ref{fig:6} are plotted for $\beta =0$. The
dashed (green) lines demarcate the boundaries of the regions with the different number
of overlapping Landau levels: the region with $N=1$ corresponds to a single contributing
level, for $N=2$ two overlapping levels contribute to the DOS.

\subsection{Analytic representations of the DOS}

It is useful to obtain the analytic representation of the DOS in terms of the $\ln \Gamma(z)$
function in the crossed fields for the Dirac Landau level case,
because it allows for the study of the DOS in the Landau level collapse regime.
Assuming, as in Eq.~(\ref{DOS-E=0-Gamma}), that all Landau levels have the same width $\Gamma$
as described by Eq.~(\ref{Lorentzian}) and the spectrum, $\mathcal{E}_n (\beta =0)$ is given by Eq.~(\ref{LL-collapse}),
the following representation for the DOS
per spin in the absence of an electric field was derived in \cite{Sharapov2004PRB}
\begin{equation}
\label{graphene-DOS-through-Gamma}
\begin{split}
& D_0(\mathcal{E},E_M) = \frac{1}{\pi^2 \hbar^2 v_F^2} \left\{
\Gamma\ln\frac{\Lambda^2}{2 E_M^2} - \right.  \\
&  \left. \mbox{Im} \left[ (\mathcal{E} + i \Gamma)
\left( \psi \left( \frac{-(\mathcal{E} + i \Gamma)^2}{2  E_M^2}\right) - \frac{E_M^2}{(\mathcal{E} + i \Gamma)^2}  \right)
\right] \right\} \\
& = \frac{1}{\pi^2 \hbar^2 v_F^2} \left[ \Gamma\ln\frac{\Lambda^2}{2 E_M^2}
 + E_M^2 \frac{d}{d \mathcal{E}} \right.\\
& \left.  {\rm Im}\left[\ln\Gamma
\left(\frac{-(\mathcal{E}+i\Gamma)^2}{2 E_M^2}\right)+
\frac{1}{2}\ln\left(\frac{-(\mathcal{E}+i\Gamma)^2}{2 E_M^2}\right)\right] \right..
\end{split}
\end{equation}
Here, $\Lambda$  denotes the energy cutoff, which is necessary due to the use of the Dirac approximation for dispersion.
The zero magnetic field, $E_M \to 0$, limit of Eq.~(\ref{graphene-DOS-through-Gamma}) reproduces the known expressions
for the DOS. In particular, using the asymptotic (\ref{digamma-expand}) we obtain \cite{Sharapov2004PRB} in the
clean limit ($\Gamma \to 0$):
\begin{equation}
\label{Dirac-free-DOS}
 D_0(\mathcal{E}) = \frac{|\mathcal{E}|}{\pi \hbar^2 v_F^2},
\end{equation}
while in the presence of impurities, we reproduce the expression from \cite{Durst2000PRB}
\begin{equation}
\label{Lee}
 D_0(0) = \frac{2 \Gamma}{\pi^2 \hbar^2 v_F^2} \ln \frac{\Lambda}{\Gamma}.
\end{equation}

It is easy to see that for the collapsing spectrum $\mathcal{E}_n (\beta)$ and the degeneracy factor $\mathfrak{g} (\beta)$,
the result can be obtained simply by replacing $E_M \to E_M (1-\beta^2)^{3/4}$.
Similarly to the nonrelativistic DOS (\ref{DOS-2DEG-Gamma}), the DOS (per unit area) in crossed fields can be expressed in terms of the DOS
(\ref{graphene-DOS-through-Gamma}) for a magnetic field only, where the representation involving the derivative with respect to the energy
$\mathcal{E}$ is particularly useful for integration
\begin{equation}
\label{graphene-DOS-through-Gamma-crossed}
D(\mathcal{E}) = \frac{1}{L_x} \int_{-L_x/2}^{L_x/2} d x_0
D_0(\mathcal{E}+ e E x_0, E_M (1-\beta^2)^{3/4}) ,
\end{equation}
where we used $x_0 = - k_y l^2$.
Then we obtain the following expression for the DOS
\begin{equation}
\label{DOS-graphene-Gamma}
\begin{split}
D(\mathcal{E}) = & \frac{1}{\pi^2 \hbar^2 v_F^2}\Gamma\ln\frac{\Lambda^2}{2 E_M^2 (1-\beta^2)^{3/2}} \\
& + \frac{2 \mathfrak{g} (\beta)}{\pi U} [d(\mathcal{E} + U/2) - d(\mathcal{E} - U/2)],
\end{split}
\end{equation}
where the function
\begin{equation}
\label{d-function}
\begin{split}
d(\mathcal{E}) =& {\rm
Im}\left[\ln\Gamma\left(
\frac{-(\mathcal{E}+i\Gamma)^2}{2 E_M^2  (1-\beta^2)^{3/2} }\right) \right.\\
& + \left.
\frac{1}{2}\ln\left(\frac{-(\mathcal{E}+i\Gamma)^2}{2  E_M^2 (1-\beta^2)^{3/2}}\right)\right].
\end{split}
\end{equation}

Similarly to the non-relativistic case, the representation in Eq.~(\ref{DOS-graphene-Gamma}) enables an analytical investigation when the
argument $z$ of the function $\Gamma(z)$ approaches infinity, using the asymptotic expansion (\ref{ln-Gamma-expand}).
While in the nonrelativistic case, this regime corresponds to the $H \to 0$ limit, for graphene,
it can be reached by either taking $E_M \to 0$ or $|\beta| \to 1$. For the case of Landau level collapse,
let us focus on the second limit.
Assuming further that the width $\Gamma \to 0$, we obtain the following result
\begin{equation}
\label{DOS-collapse}
D(\mathcal{E}) = \frac{1}{\pi \hbar^2 v_F^2}
\begin{cases}
|\mathcal{E}|, & |\mathcal{E}| > U/2,\\
\frac{\mathcal{E}^2 + U^2/4}{U}, & |\mathcal{E}| \leq U/2.
\end{cases}
\end{equation}
It can be observed that for  $|\mathcal{E}| > U/2$, the DOS matches the free
($H = E=0$) DOS of graphene, as given by Eq.~(\ref{Dirac-free-DOS}).
This occurs because the corrected degeneracy factor $\mathfrak{g}(\beta)$ eliminates the $(1-\beta^2)^{-3/2}$
divergence in the DOS that would otherwise arise \cite{Alisultanov2014JETPL,Alisultanov2014PB} (see also Ref.~\cite{Goerbig2008PRB}).

For $|\mathcal{E}| \leq U/2$, the merging of Landau levels alters the linear energy dependence of $D(\mathcal{E})$ to a parabolic one.
Additionally, the zero-energy DOS becomes finite and depends on the applied electric field, given by
$D(0) = U /(4\pi \hbar^2 v_F^2 )$.
Note that when a finite value of $\Gamma$ is considered, the DOS given by Eq.~(\ref{Lee}) must be added to Eq.~(\ref{DOS-collapse}),
showing that the product $E_M^2 (1-\beta^2)^{3/2}$ in the denominator of the first term of Eq.~(\ref{DOS-graphene-Gamma})
is properly canceled out in this limit.
This indicates that the electric field plays a similar role to $\Gamma$, making the zero energy DOS nonzero.

Equation (\ref{DOS-collapse}) can also be rewritten for $\epsilon = \mathcal{E}/E_M$, so that
\begin{equation}
\label{DOS-collapse1}
D(\epsilon) = \frac{E_M }{\pi \hbar^2 v_F^2}
\begin{cases}
|\epsilon|, & |\epsilon| > \gamma/2,\\
\frac{\epsilon^2 + \gamma^2/4}{\gamma}, & |\epsilon| \leq \gamma/2.\\
\end{cases}
\end{equation}
To conclude the discussion of Eqs.~(\ref{DOS-collapse}) and (\ref{DOS-collapse1}),
it should be noted that although these equations were derived under the conditions
$|\beta| \to 1 $ or $E_M \to 0$, they are also applicable in cases in which the ratio of electric to magnetic energy,
$\gamma$, is large and $|\mathcal{E}| \ll U$.
This is because, in such case, the argument of $\Gamma(z)$ in
Eqs.~(\ref{DOS-graphene-Gamma}) and (\ref{d-function}) also becomes significantly large.

\subsection{Numerical simulations of the DOS}

The spectrum was computed using the software ``Kwant'' \cite{Groth2014kwant}. The system is modeled as a graphene honeycomb lattice,
configured as an infinite nanoribbon with zigzag edges. The ribbon has a defined width along the $x$-direction and extends
infinitely along the $y$-direction. To explore the electronic properties, we applied an electric field along the ribbon's width,
perpendicular to its length characterized by the potential $V(\mathbf{r}) = e E x$.
In this case, it is convenient to consider the magnetic field  in the following Landau gauge $(A_{x},A_{y})=(0,H x)$, where $H$
is the magnitude of a constant magnetic field orthogonal to the ribbon's plane.
Therefore, the wave functions are plane waves in the $y$ direction $\sim \exp(i k_y y)$, where
as before $k_y$ is a continuous quantum number, which is numerically discretized for computational purposes.
By introducing the boundary conditions at the zigzag edges, we compute the energy spectrum $\mathcal{E}_{n,k_y}$.
Unlike the spectrum in Eq.~(\ref{LL-collapse}), this spectrum has a finite number of Landau levels characterized by the discrete Landau level index $n$,
due to the finite number of atoms along the $x$-direction.

Overall, the described procedure numerically implements Teller's approach \cite{Teller1931ZP} for describing electrons
in a magnetic field within a finite geometry. This approach avoids the issues inherent in Landau's method \cite{Landau1930ZP},
which uses eigenfunctions and eigenenergies derived from an infinite system to analyze the free energy in a finite geometry \cite{Hajdu1974ZP}.
Accordingly, the DOS per unit area and spin is defined as:
\begin{equation}
\label{DOS-definition-numerical}
D(\mathcal{E})= \frac{1}{L_x} \sum_{n} \int_{BZ} \frac{d k_y}{2 \pi} \delta\left( \mathcal{E}- \mathcal{E}_{n,k_y}\right),
\end{equation}
where the eigenenergies $\mathcal{E}_{n,k_y}$ are determined numerically,
the integration is performed over the first Brillouin zone in the corresponding direction, and the summation accounts for the Landau
level index $n$. This approach avoids the issues associated with Landau state counting, as discussed below Eq.~(\ref{DOS-E=0-x0}).
For the numerical calculation, the $\delta$-functions in the DOS, Eq.~(\ref{DOS-definition-numerical}),
were regularized by introducing finite energy level width in the form of a Lorentzian shape
given by Eq.~(\ref{Lorentzian}).
The width  $\Gamma$  is chosen to be much smaller than the characteristic energy scale defined by the hopping parameter $t$
and smaller than the temperature $T$ to ensure that important features remain visible.

\begin{figure}[h]
   \centering
\includegraphics[width=\columnwidth]{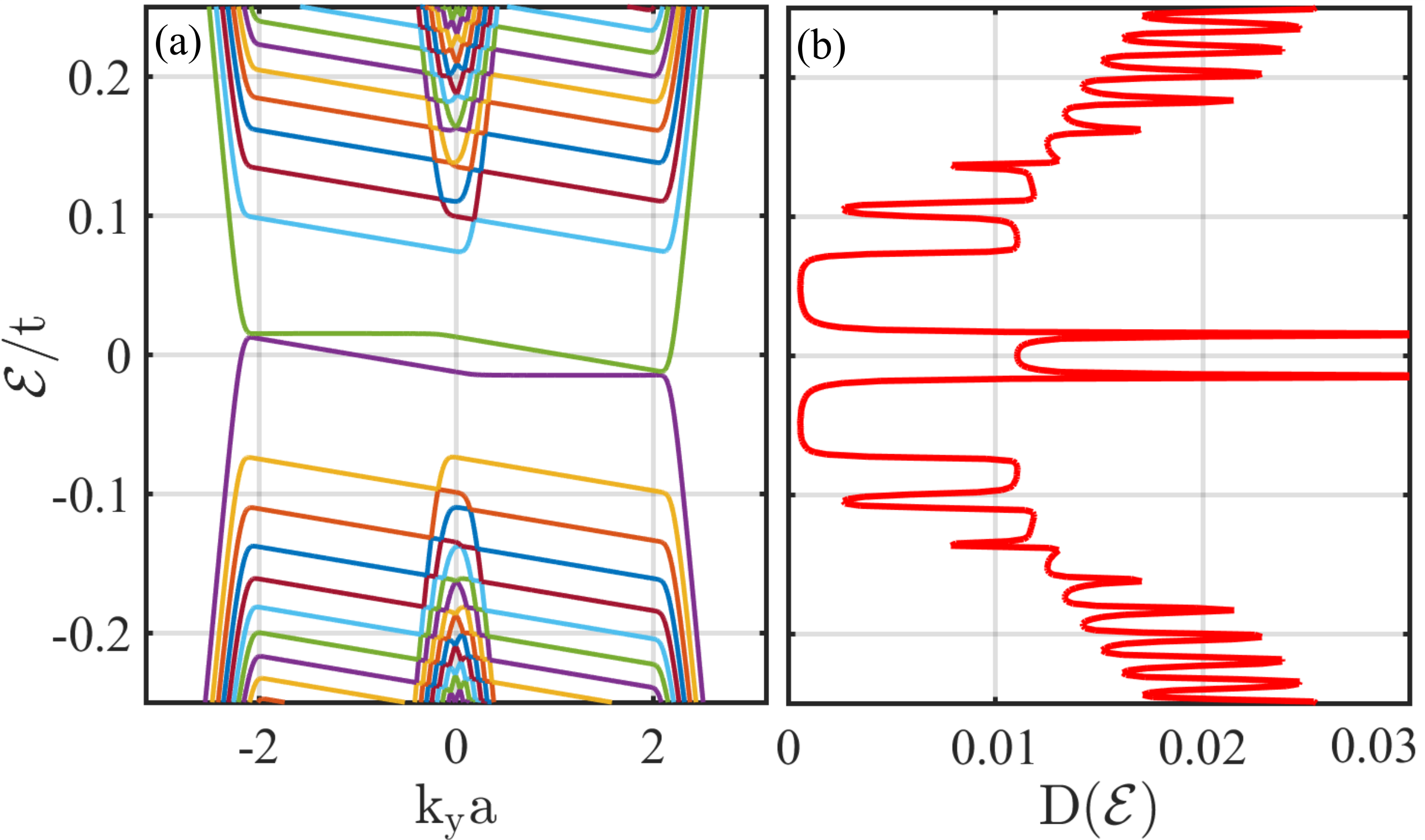}%
\caption{(a) Energy spectrum $\mathcal{E}_{n,k_y}$ computed for the tight-binding model
for a hexagonal lattice subjected to a magnetic field $H = \SI{54.5}{T}$ (or $l = 14.1 a$
where a is the triangular lattice spacing) for a system size of width $L_x = 500 a$.
The plot shows $\mathcal{E}_{n,k_y}$  in the units of the hopping parameter $t$ as function of $k_y a$, where $k_y$
is the wave vector in the $y$ direction.
An external electric field $E$ applied along $x$ axis, given
by the parameter $\beta = c E/(v_F H) = -0.014$.
(b) The DOS, $D(\mathcal{E})$, is calculated based on the depicted spectrum. The energy $\mathcal{E}$ in the units of $t$.
}
\label{fig:7}
\end{figure}

An example of such a spectrum, along with the corresponding DOS, is presented in Fig.~\ref{fig:7}.
Overall, the presented spectrum is consistent with existing lattice computations, such as those in Ref.~\cite{Lukose2007PRL}
(see also Ref.~\cite{chung2016} for a review).
Since the ribbon is sufficiently wide ($L \approx 35 l$), the spectrum corresponds to the bulk Landau levels described
by Eq.~(\ref{LL-collapse}). The electric field induces a linear $k_y$ dependence in the bulk Landau levels and
causes a constant shift in the edge states. Additionally, dispersionless states, which are surface states localized at the zigzag boundaries,
are also present. It was shown in Ref.~\cite{Herasymchuk2023PSS} by analytic methods
that these states remain unaffected by the presence of an electric field.
The energy distance between the dispersionless levels is $U$ and they show up in the DOS [Fig.~\ref{fig:7}~(b)]
as the peaks at $\mathcal{E} = \pm U/2$. The wider peaks associated with the broadening by
an electric field and starting to overlap higher Landau levels are also seen in Fig.~\ref{fig:7}~(b).

\begin{figure}[h]
   \centering
\includegraphics[width=\columnwidth]{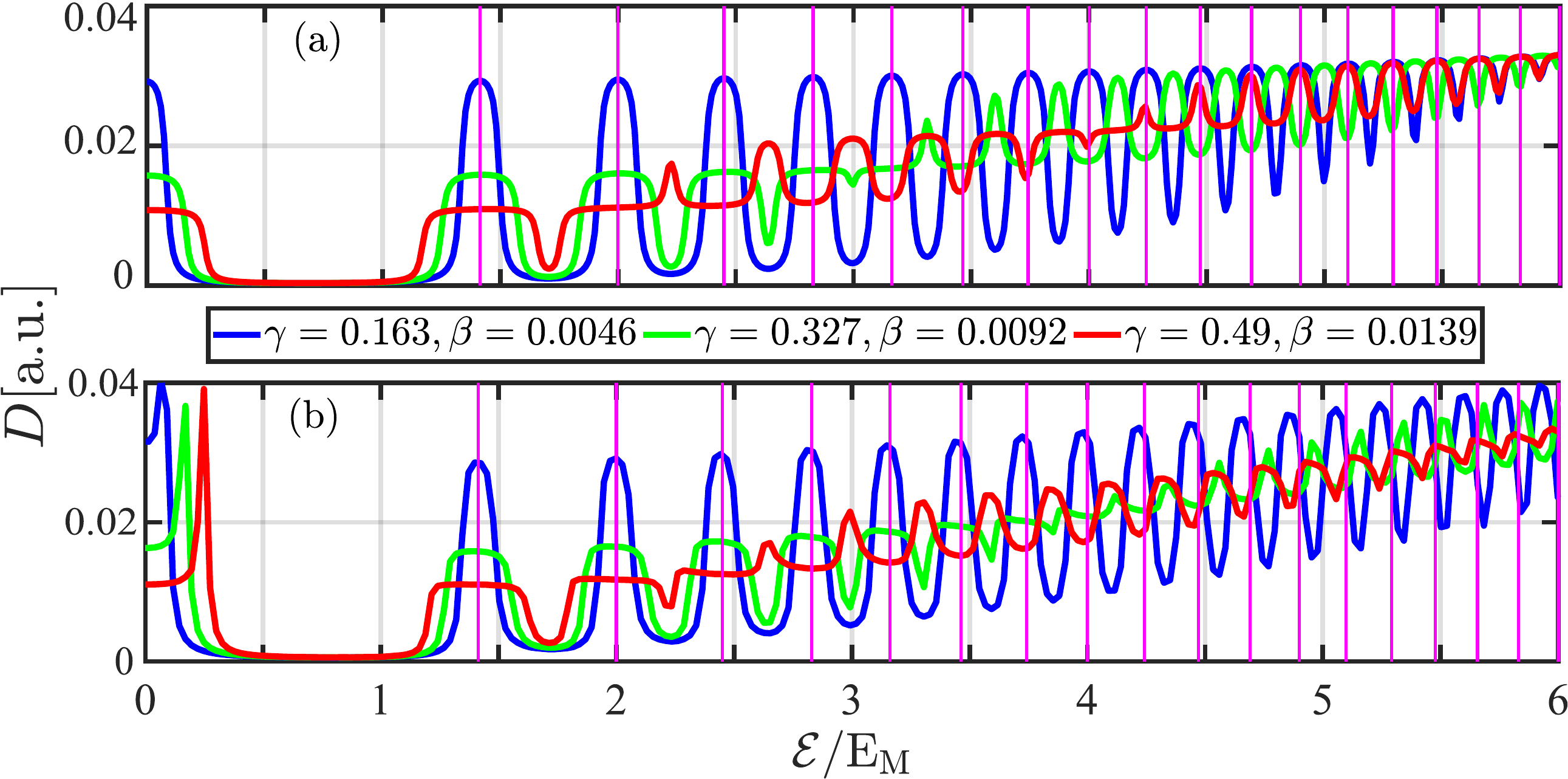}%
\caption{The DOS, $D(\mathcal{E})$, in crossed magnetic and electric fields
versus energy $\mathcal{E}$ in units of $E_M$  for $\beta = 0.0046$, $\gamma = 0.163$ (blue curve);
$\beta = 0.0092$, $\gamma = 0.327$ (green curve); and
$\beta = 0.0139$, $\gamma = 0.49$.
The magnetic field $H = \SI{54.5}{T}$, and the ribbon width equals $L_x=500 a=35.4 l$ for all curves. The scattering rate $\Gamma=0.01 t = 0.0163 E_{M}$.
The solid (magenta) vertical lines show the positions of the bulk Landau levels $\epsilon_n (\beta=0)$ given by Eq.~(\ref{LL-collapse-dimensionless}).
(a) The curves are calculated based on Eq.~(\ref{DOS-Dirac-dimensionless}).
(b) The curves are obtained by the numerical simulations on the lattice.}  \label{fig:8}
\end{figure}

To look closer at these features in Fig.~\ref{fig:8}~(a) and (b) we plot for comparison the results of the calculations
based on Eq.~(\ref{DOS-Dirac-dimensionless}) and numerical simulations, respectively, for three different values
of the electric field (or $\beta$) and the same ribbon's width $L_x=500 a \approx 35 l$.
Recall that for a fixed $L_x$, an increase in $\beta$ leads to an increase in $\gamma$, as shown in Eq.~(\ref{gamma-via-beta}).
The peaks in the blue curve, calculated for the smallest electric field, align with the positions of the bulk Landau levels in the Dirac approximation,
Eq.~(\ref{LL-collapse-dimensionless}), indicated by the solid (magenta) vertical lines.
However, even for the $n=3$ Landau level, the peak positions in Fig.\ref{fig:8}~(b), obtained from lattice calculations,
are slightly shifted from the values predicted by Eq.~(\ref{LL-collapse-dimensionless}).
This is expected since, for a strong magnetic field ($H = \SI{54.5}{T}$) , the energy of the $n=5$  level given by the Dirac approximation is
$\sqrt{5} E_M \approx 0.3 t$. This energy is high enough to deviate from the value predicted by the lattice model.
Nevertheless, for the first two levels even for the largest value of $\beta$
shown in the red curve, the agreement between the Dirac approximation and lattice simulations remains qualitative.

\begin{figure}[h]
   \centering
\includegraphics[width=\columnwidth]{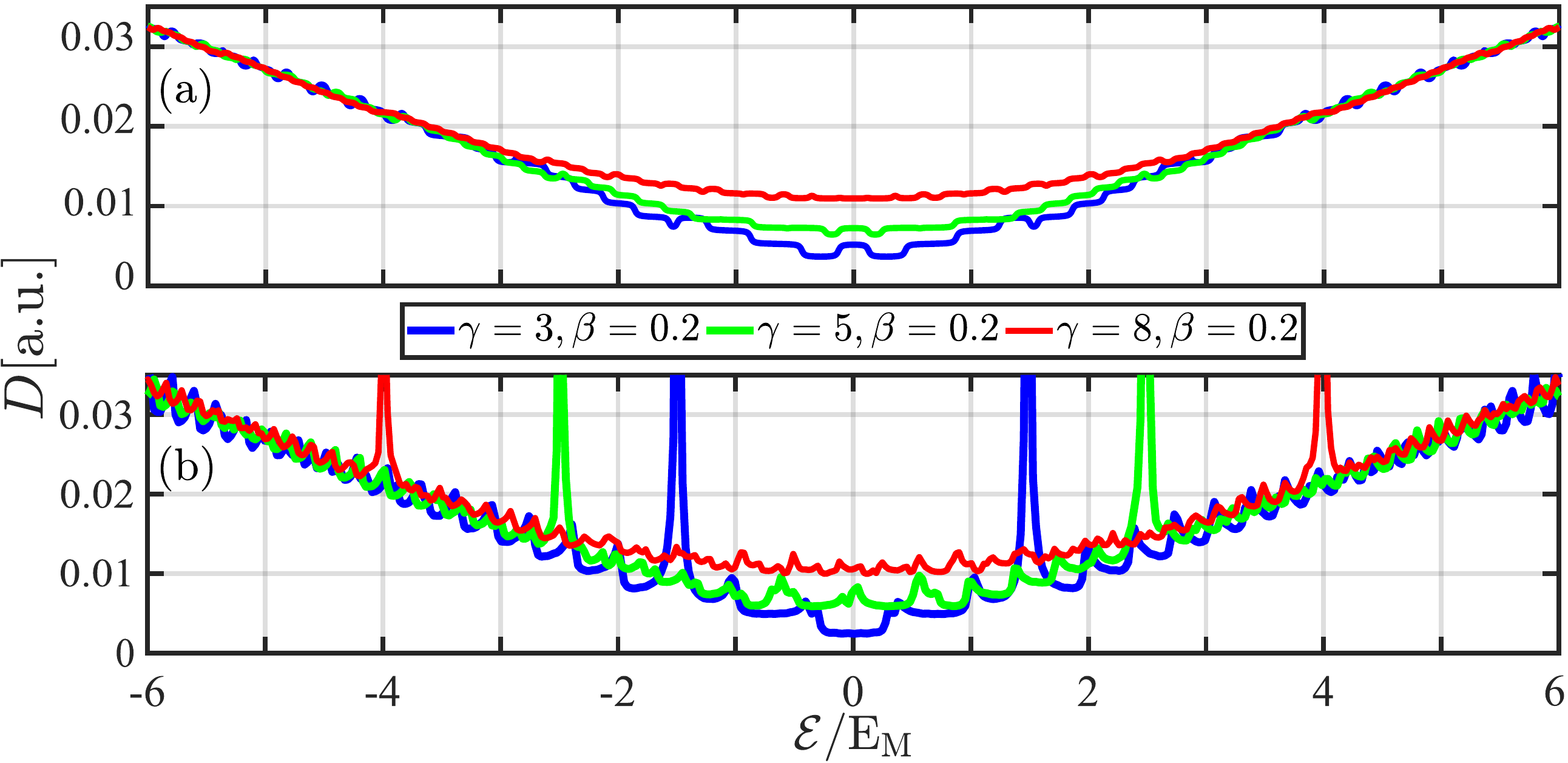}%
\caption{The DOS, $D(\mathcal{E})$, in crossed magnetic and electric fields versus energy $\mathcal{E}$ in units of $E_M$
for $\beta =0.2$ and three different values of $\gamma$:   $\gamma = 3$ (blue curve); $\gamma = 5$ (green curve);
and $\gamma = 8$ (red curve).
The magnetic field value is $H = \SI{54.5}{T}$ and the scattering rate $\Gamma=0.01t = 0.0163 E_{M}$.
(a) The curves are calculated based on Eq.~(\ref{DOS-Dirac-dimensionless}).
(b) The curves are obtained by the numerical simulations on the lattice.}  \label{fig:9}
\end{figure}

Now, using simulations, we test the Landau level degeneracy factor $\mathfrak{g} (\beta)$ introduced in Eq.~(\ref{degeneracy-Dirac}),
which depends on $\beta$.
In Figs.~\ref{fig:9}~(a) and (b) we plot for comparison  the results of the calculations
based on Eq.~(\ref{DOS-Dirac-dimensionless}) and numerical simulations, respectively, for three different values
of $\gamma$ and the fixed $\beta$. In accordance with Eq.~(\ref{gamma-via-beta}) the variation
of $\gamma$ is reached by changing the ribbon's width $L_x$.
The peaks associated with the dispersionless levels at $\pm \gamma/2$ are visible in Fig.~\ref{fig:9}~(b)
but are absent  in Fig.~\ref{fig:9}(a), which was computed considering only the bulk Landau levels.
However, except for these peaks, there is good agreement between the results obtained from
Eq.~(\ref{DOS-Dirac-dimensionless}) and numerical simulations.
The DOS between the peaks is parabolic and described by Eq.~(\ref{DOS-collapse1}).

\begin{figure}[h]
   \centering
\includegraphics[width=\columnwidth]{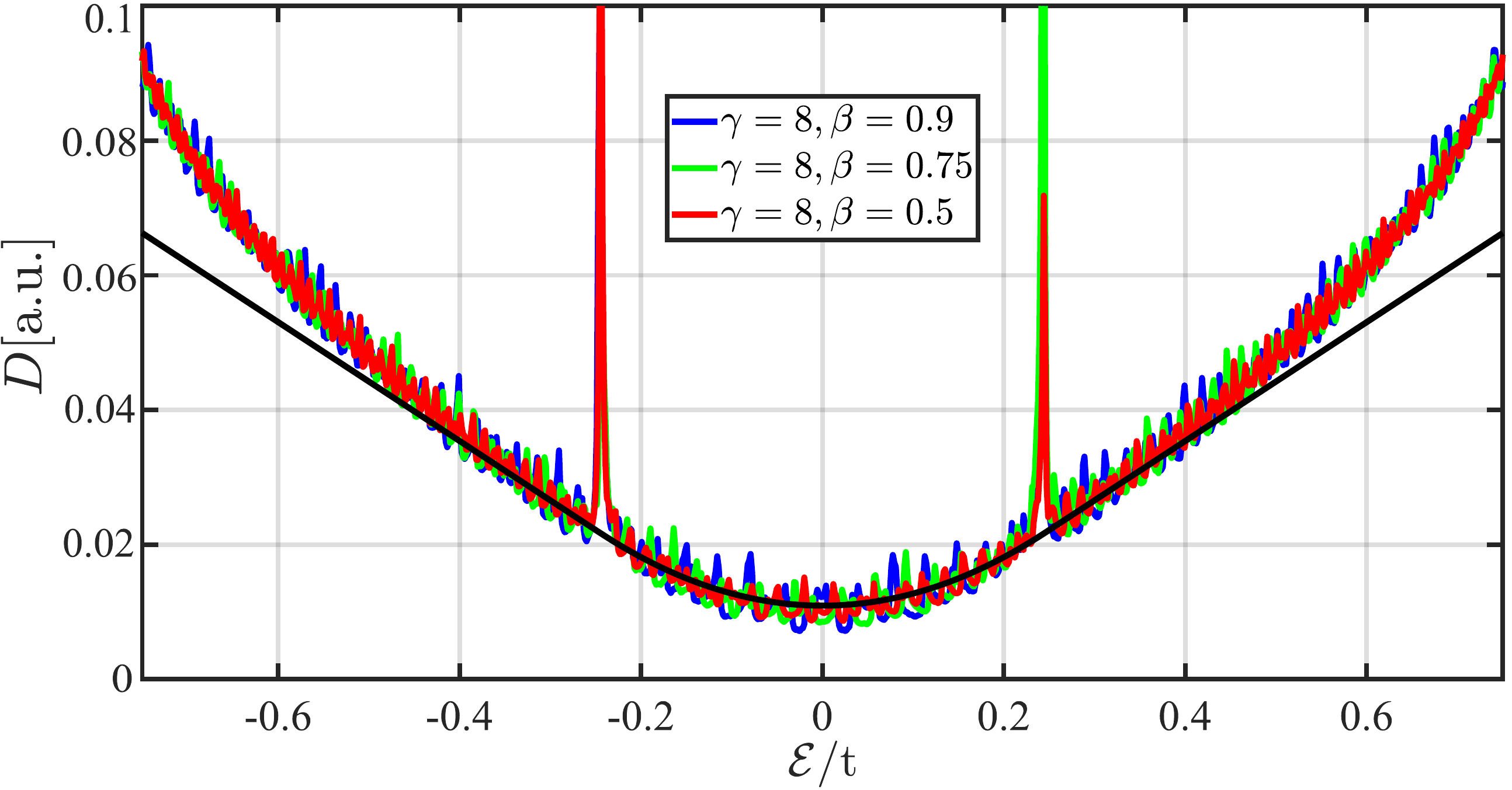}%
\caption{The DOS, $D(\mathcal{E})$, in crossed magnetic and electric fields
versus energy $\mathcal{E}$ in units of  the hopping parameter $t$  by numerical simulations
for $\gamma =8$ and three different values of $\beta$ : $\beta = 0.9$ (blue curve);
$\beta = 0.75$ (green curve); and $\beta = 0.5$ (red curve).
The magnetic field value is $H = \SI{54.5}{T}$ and the scattering rate $\Gamma=0.001t = 0.0163 E_{M}$.
The black curve is the DOS given by Eq.~(\ref{DOS-collapse1}).
}  \label{fig:10}
\end{figure}

To see this in Fig.~\ref{fig:10} we plot together the DOS given by
Eq.~(\ref{DOS-collapse1}) (black curve) and numerical simulations, respectively, for three different values
of $\beta$ and the fixed $\gamma$.
In accordance with Eq.~(\ref{gamma-via-beta}) the variation
of $\beta$ is reached by changing the ribbon's width $L_x$.
We observe that the simulation results for different values of $\beta$ are nearly identical and
align well with Eq.~(\ref{DOS-collapse1}). It is important to emphasize that the DOS described by Eq.~(\ref{DOS-collapse1})
is independent of $\beta$. However, if the factor $(1-\beta^2)^{3/2}$ were not included in the degeneracy factor given
by Eq.~(\ref{degeneracy-Dirac}), the three theoretical curves corresponding to different values of $\beta$ would deviate significantly.
Such discrepancies would contradict the numerical simulations.

For $|\epsilon| \geq \gamma/2$ the parabolic behavior observed in the numerical simulations transforms to a linear one,
in agreement with Eq.~(\ref{DOS-collapse1}). However, as $|\epsilon|$ continues to increase, the linear behavior in the
numerical curves becomes nonlinear once again. This occurs for
$|\epsilon| \gtrsim  0.3 t$, where the Dirac approximation is no longer applicable.

\section{Differential entropy}
\label{sec:entropy}

As mentioned in the Introduction, the differential entropy $s$, also known as the entropy per particle, is directly related to
the temperature derivative of the chemical potential at a fixed electron density $n(\mu, T)$ [see Eq.~(\ref{entropy-part})].
The latter can be obtained using the thermodynamic
identity
\begin{equation}
\label{derivative}
\left( \frac{\partial \mu}{\partial T} \right)_n = - \left( \frac{\partial n}{\partial T} \right)_\mu
\left( \frac{\partial n}{\partial \mu} \right)_T^{-1}.
\end{equation}
At thermal equilibrium, the total density of electrons is
\begin{equation}
\label{number}
n (T,\mu)  =\int _{-\infty}^{\infty }d \epsilon D (\epsilon ) f_{\rm{FD}} \left(\frac{\epsilon -\mu }{T} \right),
\end{equation}
where $f_{\rm{FD}} ( x ) = 1/[\exp(x)+1 ]$ is the Fermi-Dirac distribution function
and  we set the Boltzmann constant to $k_B=1$ and measure
the temperature in energy units.
Note that in the presence of electron-hole symmetry, it is convenient to work with the difference between the electron
and hole densities rather than the total electron density, as is commonly done for graphene \cite{Tsaran2017SciRep}.

Differentiating Eq.~(\ref{number}) with respect to $T$ and $\mu$,
respectively, results in the well-known expression for the
differential entropy \cite{Varlamov2016PRB,Tsaran2017SciRep}
\begin{equation} \label{entropy-part-DOS}
s (\mu,T)
=\frac1T\frac{\int _{-\infty}^{\infty }d \varepsilon\,  D (\varepsilon ) (\varepsilon - \mu) \cosh^{-2}
\left( \frac{\varepsilon - \mu}{2T} \right) }{\int _{-\infty}^{\infty }d \varepsilon D (\varepsilon ) \,
\cosh^{-2} \left( \frac{\varepsilon - \mu}{2T} \right) }.
\end{equation}
Using Eq.~(\ref{entropy-part-DOS}), it is clear that the extrema in the dependence of $D(\mu)$ correspond to the zeros of $s(\mu)$.
Since van Hove singularities produce sharp peaks in the DOS, they appear as characteristic features in the dependence
of $s(\mu)$ discussed in \cite{Kulynych2022}, followed by pronounced peak and dip structures
\cite{Varlamov2016PRB, Tsaran2017SciRep}.

There is a similarity between Eq.~(\ref{entropy-part-DOS}) and the corresponding expression for the Seebeck coefficient $\mathcal{S}$,
such that,  for an energy-independent relaxation time, they are identical.
The relationship between $\mathcal{S}$ and $s$ in zigzag graphene ribbons without external fields was analyzed in detail
in Ref.~\onlinecite{Cortes2023PRB}, where it was shown that within the gap, $\mathcal{S} \simeq s/e$, with $e$ as the electron charge
and the Boltzmann constant restored in $s$.

\begin{figure}[h]
   \centering
\includegraphics[width=\columnwidth]{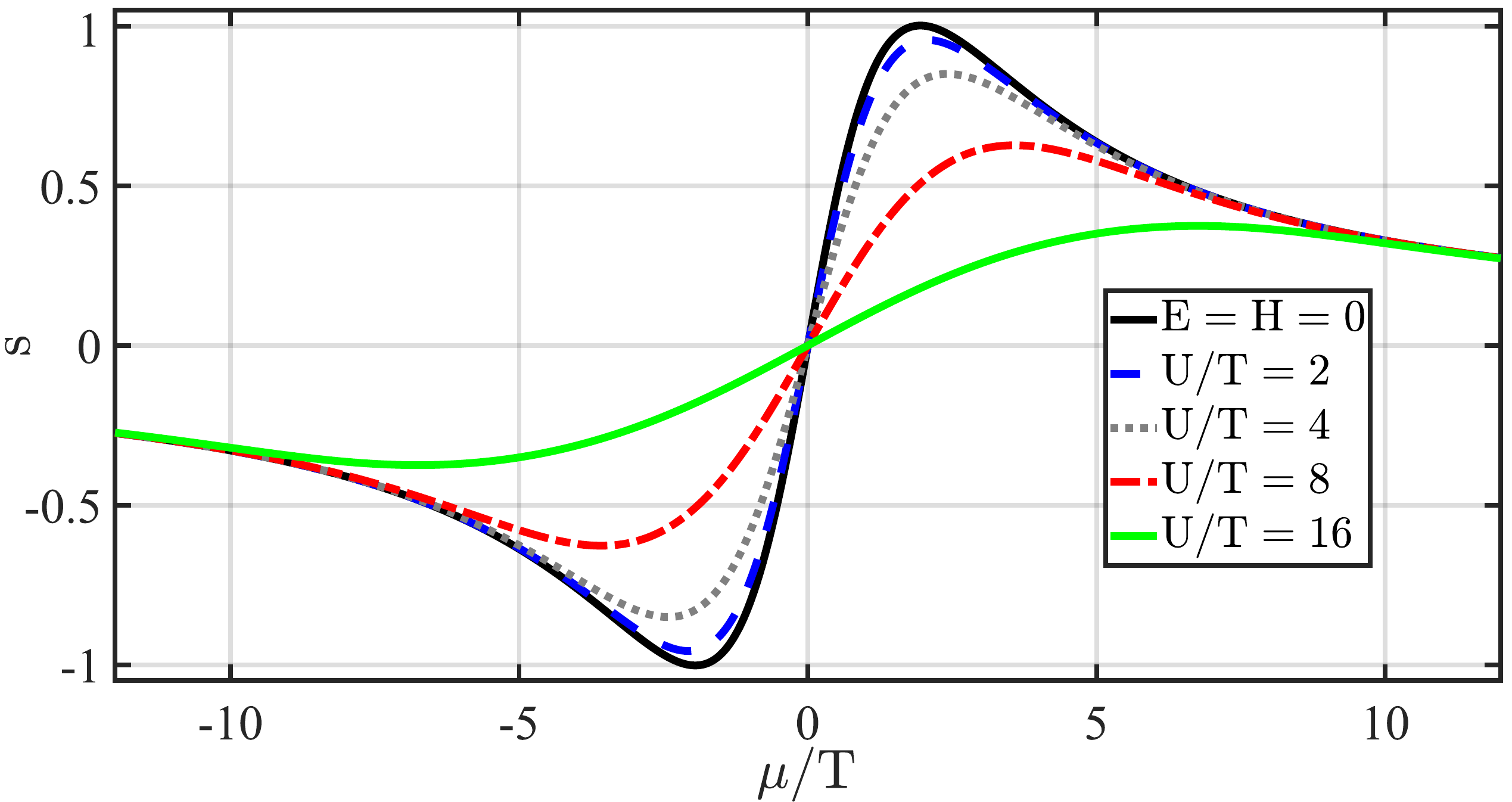}%
\caption{The differential entropy $s$ versus the chemical potential $\mu$ in
units of $T$. The solid (black) curve is for the free $H= E=0$ case; the dashed (blue),
$U = 2 T$;  the dotted (grey), $U = 4 T$;  the dash-dotted (red),  $U = 8 T$; thick solid
(green), $U = 16 T$.
}
\label{fig:11}
\end{figure}

In Fig.~\ref{fig:11} we plot differential entropy as a function of the chemical potential
$\mu$ for the bulk states only.
The reference case $H = E =0$, where the free DOS is given by Eq.~(\ref{Dirac-free-DOS}),
is represented by the solid black curve and is analytically described by the following expression \cite{Tsaran2017SciRep}
\begin{equation}
\label{derivative-Delta=0}
\begin{split}
s & = \frac{1}{\ln \left (2 \cosh \frac{\mu}{2T} \right)}
\left[ \mbox{Li}_2 \left(-e^{- \frac{\mu}{T}} \right) - \mbox{Li}_2 \left(-e^{ \frac{\mu}{T}}\right) \right]
-\frac{\mu}{T}.
\end{split}
\end{equation}
Here $\mbox{Li}(z)$ is the polylogarithm function. One can also extract the asymptotics of Eq.~(\ref{derivative-Delta=0}) both
for $|\mu| \ll T$ and $|\mu| \gg T$ \cite{Tsaran2017SciRep}.
In particular, when multiplied by the factor $k_B/e$, the latter limit matches the Seebeck coefficient for a free-electron gas.
The other curves in Fig.~\ref{fig:11} were computed using the DOS (\ref{DOS-collapse}) in the collapse limit
and for large values of $\gamma$.
As the value of $U$ given in units of temperature $T$ increases, these
curves become less steep, reflecting a more smooth behavior of the DOS.
It should also be noted that the differential entropy is independent of whether the DOS is calculated using the
degeneracy factor $\mathfrak{g}(\beta)$ or $\mathfrak{g}_L$.

\begin{figure}[h]
   \centering
\includegraphics[width=\columnwidth]{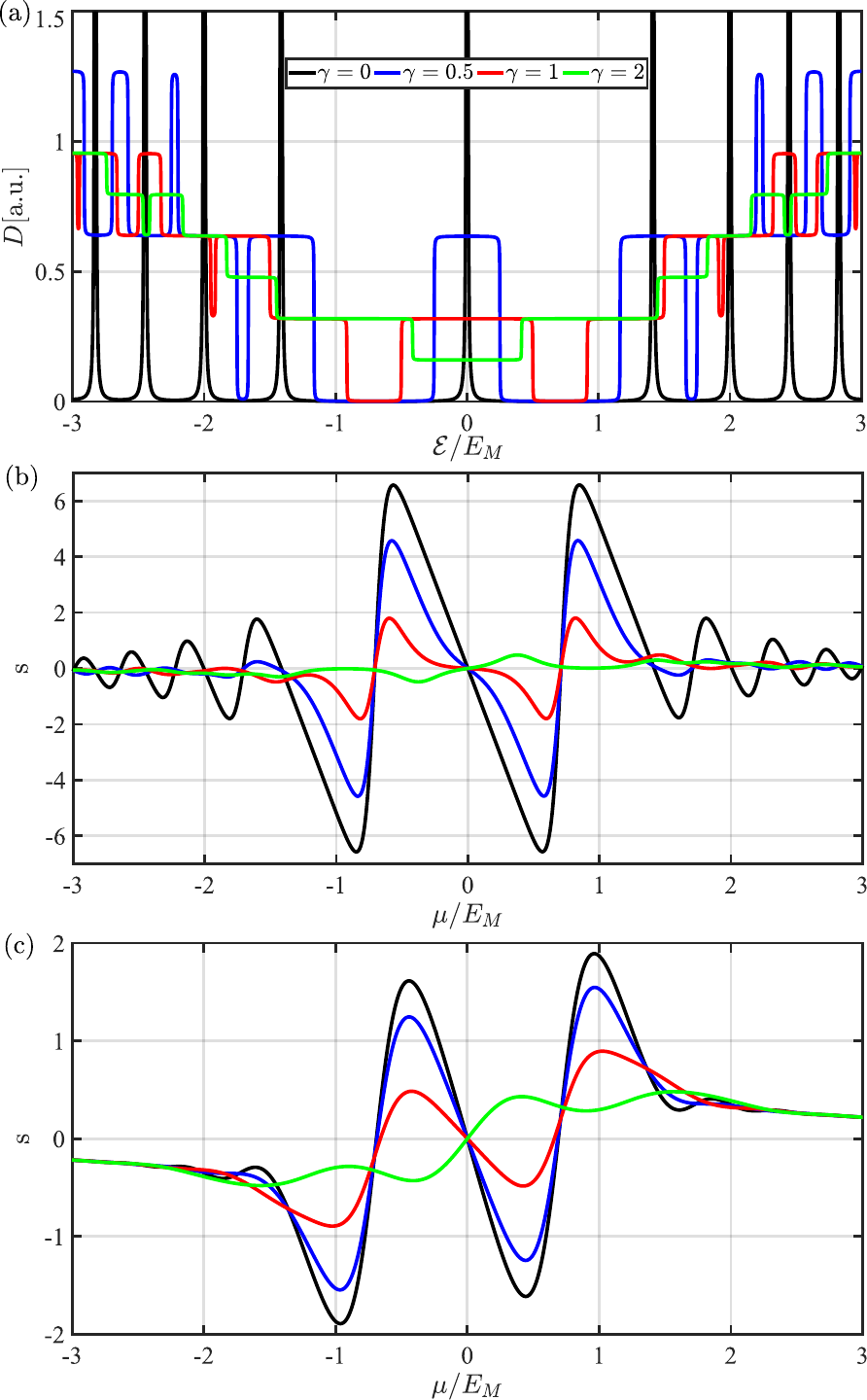}%
\caption{The DOS, $D(\mathcal{E})$, and differential entropy $s(\mu)$ for four values of $\gamma$:
$\gamma =0$ (black curve); $\gamma =0.5$ (blue curve); $\gamma =1$ (red curve); and $\gamma =2$ (green curve).
(a) The DOS, $D(\mathcal{E})$ versus energy $\mathcal{E}$ in
units of $E_M$. (b) The differential entropy $s$ versus the chemical potential $\mu$ in the
units of $E_M$, for $T=0.08 E_M$ (c) The same as (b), but for for $T =0.2 E_M$.
}
\label{fig:12}
\end{figure}

Next, we analyze the behavior of $s(\mu)$ for small values of $\gamma$, where the lowest Landau levels remain distinct and do not overlap.
In Fig.~\ref{fig:12} we plot the DOS and the corresponding
differential entropy $s$ as a function of the chemical potential at two values of the temperature $T=0.08 E_M$ and $T=0.2 E_M$
in four cases: $\gamma =0$ ($E =0$), $\gamma =0.5$, $\gamma = 1$ and $\gamma =2$.
The DOS, $D(\mathcal{E})$ in  Fig.~\ref{fig:12}~(a) is computed using Eq.~(\ref{DOS-Dirac})
[see also Eq.~(\ref{DOS-Dirac-dimensionless})]. As in Sec.~\ref{sec:DOS-Dirac}
for convenience of comparison, we set $\beta =0$. The case marked as $\gamma=0$ corresponds
to the $\gamma \to 0$ limit when the results in graphene under a magnetic field alone
are recovered. The value $\Gamma = 0.001 E_m$ is taken.
The  dependence $s(\mu)$ exhibits oscillations and features a sharp peak when the chemical potential is near the Dirac point,
$|\mu| \sim T$,  within the temperature range. The amplitude of the oscillations is higher for the lower value of $T=0.08 E_M$.
We observe that as the value of $\gamma$ increases, these features become less pronounced.
Note that for $\gamma =2$, when the $n=0$ and $n=1$ levels begin to overlap, the corresponding green curve crosses the $\mu =0$
point from negative to positive values of $s$. In contrast, curves with smaller values of $\gamma$ cross this point from positive
to negative values of $s$.

\begin{figure}[h]
   \centering
\includegraphics[width=\columnwidth]{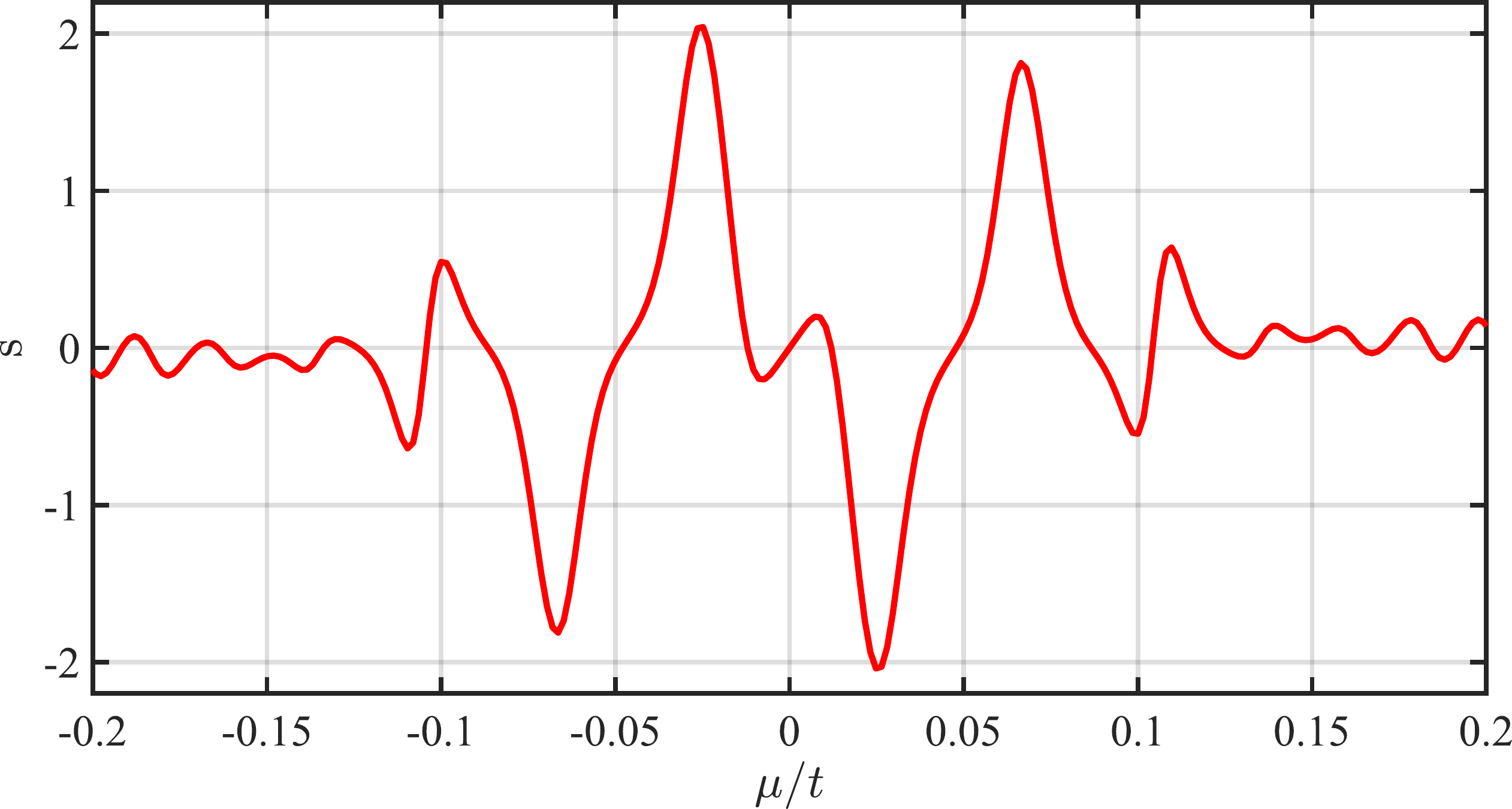}%
\caption{The differential entropy $s$ versus the chemical potential $\mu$ in
units of $t$, for $T=0.003t = 0.049 E_{M}$ and $U=0.03t$, with scattering rate $\Gamma=0.001t = 0.0163 E_{M}$,
calculated by numerical simulations on the lattice.
}
\label{fig:13}
\end{figure}

Finally, in Fig.~\ref{fig:13} we show the dependence $s(\mu)$ computed for the DOS presented in
Fig.~\ref{fig:7}~(b).
Overall the behavior of $s(\mu)$ is similar to the results presented in  Fig.~\ref{fig:12}, with one significant exception.
The DOS peaks at $\mathcal{E} = \pm U/2 \approx \pm  0.015 t$,
associated with the dispersionless surface states, introduce an additional dip-peak structure near
$\mu =0$.

\section{Conclusion}
\label{sec:concl}

One can estimate that for a magnetic field of
$H =\SI{1}{T}$ and the Fermi velocity $v_F = \SI{1e6}{m/s}$,
the critical electric field
is $E_c = \SI{1e4}{V/cm}$, which is achievable experimentally.
For example, with $H =\SI{0.1}{T}$, the magnetic length is approximately $l \approx \SI{82}{\nm}$.
For a ribbon of width $L_x = 20 l \approx \SI{1.6}{\um}$,
this corresponds to a critical voltage between the ribbon edges of $U = \SI{0.16}{V}$.
Indeed, experimental studies have reported the realization of Landau level collapse
\cite{Singh2009PRB, Gu2011PRL}.
Another potential approach to achieve Landau level collapse is by generating strain-induced pseudomagnetic or electric fields, as proposed in
Refs.~\cite{Castro2017PRB, Grassano2020PRB}.
This demonstrates that the results for the DOS and differential entropy presented in this work are accessible
for experimental investigation.

It would be particularly interesting to explore how the behavior of the DOS can be adjusted by applying an
in-plane electric field, given its relevance from both research and practical perspectives. From a research
point of view, it is crucial to confirm that the Landau degeneracy factor
$\mathfrak{g}_L$ [see Eq.~(\ref{degeneracy-LL})]
for graphene in the crossed fields must be replaced with the electric field-dependent factor
$\mathfrak{g} (\beta)$, as defined in Eq.~(\ref{degeneracy-Dirac}).
It would also be useful to study
the degeneracy factor in other similar situations.
From a practical perspective, enhancing the DOS by tuning the in-plane electric field could be valuable for achieving
greater control over the transport properties of graphene.

Using the DOS results, we analyzed the behavior of the differential entropy. As the energy dependence of the DOS for bulk states becomes smoother, the differential entropy varies less steeply compared to the $H = E = 0$ case.
When the lowest Landau levels do not overlap,  the  dependence $s(\mu)$ exhibits oscillations and features a sharp peak when the chemical
potential is near the Dirac point.
Further investigation is warranted to understand how the dip-peak structure observed in  $s(\mu)$ near $\mu =0$,
associated with the dispersionless surface states, manifests in the Seebeck coefficient.

\begin{acknowledgments}

We would like to thank  V.P.~Gusynin, A.A.~Herasymchuk, and A.A.~Varlamov for stimulating discussions.
We would like to express our gratitude to M.O.~Goerbig for insightful comments.
We are grateful to the Armed Forces of Ukraine for providing security to perform this work.
D.O.O. acknowledges the support by the Kavli
Foundation. S.G.Sh. acknowledges support from the National Research Foundation of Ukraine grant  (2023.03/0097)
``Electronic and transport properties of Dirac materials and
Josephson junctions.'' The authors acknowledge the use of computational resources of the DelftBlue supercomputer,
provided by Delft High Performance Computing Centre \cite{DHPC2024}.

\end{acknowledgments}

\bibliography{LL-collapse-entropy.bib}

\end{document}